\documentclass[a4paper,twoside]{article}
\usepackage{multirow}
\newcommand{\affil}[1]{$^{\rm #1}$}
\usepackage[authoryear]{natbib}
\bibpunct{(}{)}{;}{a}{}{,}
\usepackage{graphicx}
\date{} %Please leave the date blank
\newcommand{\etal}{\emph{et\,al.}}
\title{\large\bf Optimal Extraction of Fibre Optic Spectroscopy}
\author{\parbox{\textwidth}{\flushleft
\vspace{-0.5cm}
{\it R. Sharp\affil{A,B}, M.N. Birchall\affil{A}}\\
\vspace{0.4cm}
{\small \affil{A}\,Anglo-Australian Observatory, PO Box 296, Epping, NSW, 1710, Australia}\\
{\small \affil{B}\,Email: rgs@aao.gov.au}}}

\begin{document}
\twocolumn[
\begin{changemargin}{.8cm}{.5cm}
\begin{minipage}{.9\textwidth}
\vspace{-1cm}
\maketitle
%
%
%%%%%%%%%%%%%     ABSTRACT    %%%%%%%%%%%%%
%Abstract of no more than 200 words here.
\small{\bf Abstract:} We report an \emph{optimal extraction}
methodology, for the reduction of multi-object fibre spectroscopy
data, operating in the regime of tightly packed (and hence
significantly overlapping) fibre profiles.  The routine minimises
crosstalk between adjacent fibres and statistically weights the
extraction to reduce noise.  As an example of the process we use
simulations of the numerous modes of operation of the AAOmega fibre
spectrograph and observational data from the SPIRAL Integral Field
Unit at the Anglo-Australian Telescope.

%%%%%%%%%%%%%     KEYWORDS    %%%%%%%%%%%%%
\medskip{\bf Keywords:} instrumentation: spectrograph, methods: data analysis
% Please write all keywords in lower case. PASA uses the
% standard list of subject headings adopted by The Astrophysical Journal
% and available from http://www.journals.uchicago.edu/ApJ/keywords_text.html.
% Keywords are separated by em-dashes, i.e. ---

%%%%%%%%DO NOT EDIT%%%%%%%%%%%%
\medskip
\medskip
\end{minipage}
\end{changemargin}
]
\small
%%%%%%%%EDIT FROM HERE%%%%%%%%%%%%

\section{Introduction}
Fibre-optic multi-object spectroscopy is a powerful tool with which to
segment the focal plane of a wide-field spectroscopic telescope.  The
technique, pioneered in the 1970s (see \citet{hill1988} for a
historical review), has been repeatedly put to use for major
undertakings in astronomy.  The high multiplex advantage that can be
achieved, typically over wide fields-of-view, using fibre optic feeds
easily outweighs the expense of implementation and inherent
limitations of a fibre feed system for many classes of astronomical
observation.  Notable recent examples of include the 2dF Galaxy
Redshift Survey \citep{col2001}, 2dF Quasar survey \citep{boy2000} and
the Sloan Digital Sky Survey \citep{york2000}.

The use of coherent fibre bundles to record two-dimensional spatially
resolved information at the spectrograph slit is an obvious extension
of the technology, and indeed a number of the current generation of
Integral Field Spectrograph (IFS) systems use fibre feeds to reformat
the focal plane (e.g.\ AAT-SPIRAL, Gemini-GMOS, ESO-VIMOS \& ARGUS).

However, increased multiplex (or field of view for IFS instruments)
comes at a price.  With CCD (and IR array) pixels still at a premium
in any astronomical instrument, there is a need to tightly pack the
fibres into the spectrograph slit, minimising the \emph{dead space}
between adjacent fibres in order to maximise the number of fibres
available for a given instrument format.  But tight fibre packing
means poor sampling of individual fibre Point Spread Functions (PSF)
and can result in significant overlap of the profiles of adjacent
fibres leading to strong cross-contamination, or {\it cross-talk},
between fibres.  Recovering the spectral information for spectra
observed in tightly packed systems is therefore complex.  Simple
summation of the pixel values surrounding the peak of a fibres' trace
is doomed to failure due to contamination from adjacent fibres
(particularly with high contrast observations as are common in IFS
applications).  Even if individual fibres are \emph{reasonably} well
resolved, the extraction will often suffer increased noise from
contamination by the wings of adjacent fibre profiles, or poor
weighting of pixel values introducing increased CCD read-noise in the
limit of low background observations.

What is required is an extraction procedure which takes account of the
interaction between adjacent fibres and provides a statistically
\emph{optimal} estimate of the true intensity for each fibre spectrum.
We have recently implemented such an extraction mechanism for the
AAOmega spectrograph system at the Anglo-Australia Telescope.  The
code has been developed in the \texttt{IDL} programing language and
implementation within the \texttt{2dfdr} data reduction environment
(constructed primarily in \texttt{FORTRAN}).  In this work we quantify
the need for such an extraction algorithm and demonstrate that while
there is limited gain in using the routine for the well resolved fibre
profiles of the default AAOmega instrument Multi-Object Spectroscopy
(MOS) mode, it is essential for high purity data taken with the tight
fibre packing utilised for the recently implemented AAOmega
mini-shuffling {\it nod-and-shuffle} observing mode and for the SPIRAL
IFS feed to the AAOmega spectrograph.  We describe the methodology
used and demonstrate its application to observations taken with the
AAOmega-SPIRAL system.

The paper is organised as follows.  In \S~\ref{spec ext} \&
\ref{instrument modes} we introduce the three alternative spectra
extraction methodologies currently implemented with in the
\texttt{2dfdr} data reduction software and four modes of instrument
operation for the AAOmega spectrograph system at the AAT. In
\S~\ref{need for crosstalk correction} we demonstrate the need for an
extraction algorithm which accounts for fibre-to-fibre cross-talk.
The two non-trivial extraction algorithm from \S~\ref{spec ext} are
then described in detail in \S~\ref{algorithams}.  The procedure for
applying the the {\it optimal extraction algorithm} is given in
\S~\ref{proc} \& \ref{non iter BG}.  Scattered light effects are
addressed in \S~\ref{scattered light} and example data from the
AAOmega-SPIRAL system is presented \S~\ref{discussion}.

\section{Spectral extraction methodologies}
\label{spec ext}
Three spectral extraction methodologies are currently implemented
within the {\tt 2dfdr} data reduction environment used with data from
the AAOmega spectrograph.  In each case the starting point is a map of
the fibre profile centroids on the CCD.  Since this fibre map is a
series of near parallel tracks it has historically been labeled the
{\it tramline map}.  The generation of the tramline map is conducted
separately from the spectral extraction and is not discussed in detail
here.  Briefly, the fibre profile centroids are determined at
intervals across the CCD (every $\sim$50 pixels along the dispersion
axis) and a low order polynomial model is fitted to the centroids for
each fibre, guided by an optical model for the expected spectrograph
camera distortions.

The three spectral extraction methods currently available within {\tt
2dfdr} are:\\

\vspace{-0.25cm}\noindent{\it Tramline summation -} This most basic of
extractions is obtained by simple summation of all pixel values
associated with a given fibre.  The spectra for the fibres are
dispersed broadly along rows of CCD pixels and each column of the CCD
is treated independently.  The summation range for each fibre typically
runs over the pixels bounded by the mid-points between the two
adjacent fibre profiles.  While quick to compute, the simple tramline
extraction propagates the maximum CCD readout-noise into the final
extraction since it gives equal weight to all pixels in the summation
regardless of the flux level of the fibre profile in a given pixel.
Tramline summation also suffers an aperture loss effect if the
inter-fibre gap over which pixels are summed is not significantly
larger than the fibre profile width. This {\it aperture correction}
can of course be accounted for if the true fibre profile is known.\\

\vspace{-0.25cm}\noindent{\it Gaussian weighted summation -} An
algorithm for performing a weighted Gaussian summation over a single
isolated fibre profile (via a Least Squares Fit) is discussed in
section \ref{Gaussian extraction}.  This weighted summation minimises
the contribution for CCD readout noise and does not suffer an aperture
effect.  This mode is the default for AAOmega-MOS spectroscopy and we
show in \S\ref{need for crosstalk correction} that for well separated
spectra it delivers acceptable results.\\

\vspace{-0.25cm}\noindent{\it Multi-fibre deconvolution -} The ideal
solution for the minimisation of cross-talk between fibres is an
extraction algorithm which performs a multi-fibre deconvolution of the
data given an underlying model assumption for the fibre profiles on
the CCD.

The extent to which fibre-to-fibre cross talk is present under each
extraction models described above is tested in \S\ref{need for
crosstalk correction}.  The algorithms adopted for the {\it Gaussian}
and {\it multi-fibre deconvolution} solutions are then discussed in
\S\ref{Gaussian extraction} and \S\ref{the algoritham}.  All three
approaches have been implemented for the AAOmega spectrograph at the
Anglo-Australia Telescope.  The multiple operating modes of AAOmega
are outlined in the next section.

\section{The fibre spectrograph instrument modes}
\label{instrument modes}
Table~\ref{Crosstalk modes} outlines four instrument modes for the
AAOmega facility at the Anglo-Australia Telescope
\citep{sau2004,sha2006}.  Data from the AAOmega system is primarily
processed using the \texttt{2dfdr} data reduction software and the
three extraction algorithms presented above have been developed from
the code base of the previous 2dF spectrographs.

\subsubsection*{AAOmega MOS}
The first instrument mode is the default multi-object spectroscopy
(MOS) mode of operation.  In this mode 392 science fibres are deployed
on astronomical targets across the $\pi$\,deg$^2$ field-of-view of the
2dF prime focus corrector.  The fibres feed the dual beam AAOmega
spectrograph, each arm of which is equipped with a 2k$\times$4k E2V
CCD.  This results in a fibre-to-fibre separation, the fibre {\it
pitch}, of $\sim$10 pixels.  The cameras deliver largely Gaussian
profiles of FWHM$\sim$3.4 CCD pixels.

\subsubsection*{Mini-Shuffling}
The second mode under consideration is the newly implemented {\it
mini-shuffling} mode.  The high quality AAOmega PSF makes it possible
to undertake fibre {\it nod-and-shuffle} observations for high quality
sky subtraction \citep{gla2001} by interleaving multiple on-sky
exposures on a single CCD frame using the inter-fibre gaps to provide
the required CCD storage areas during charge shuffling.  This mode
effectively doubles the number of AAOmega fibres on a CCD by halving
the fibre pitch.  This introduces significant fibre-to-fibre crosstalk
as will be demonstrated in \S\ref{need for crosstalk correction}.  A
subsequent paper will discuss the effectiveness of {\it
mini-shuffling} in detail.
% \citep{sharp}.

\subsubsection*{The SPIRAL Integral Field Unit}
The optimal extraction algorithm presented in \S\ref{the algoritham}
was developed for the AAOmega-SPIRAL Integral Field Spectrograph
system used with AAOmega at the AAT.  It is based on an implementation
developed for the CIRPASS IR spectrograph \citep{par2004}.  Both
CIRPASS and SPIRAL use a tight fibre packing, and with good reason.
CIRPASS initially utilised a Rockwell Hawaii-1K IR array and hence
required a tight fibre packing to maintain the wide field of view in
IFS mode \citep{kra2006}.  The SPIRAL system is designed to allow
\emph{nod-and-shuffle} \citep{gla2001} observations with the full
22.4$\times$11.2\,arcsec FoV of the SPIRAL system, while using the
same detector real-estate as the AAOmega multi-object fibre feed from
the 2dF fibre positioner (Sharp \etal\ 2006).  Hence for the SPIRAL
system CCD pixels are at a premium as each fibre essentially requires
twice as many pixels on the CCD as dictated by its intrinsic foot
print.

\subsubsection*{HERMES}
The final mode of operation considered in Table~\ref{Crosstalk modes}
is that proposed for the HERMES high resolution (R$\sim$30,000)
multi-object stellar spectrograph under development for use with the
2dF positioner at the AAT.  HERMES will employ 4k$\times$4k CCDs with
15$\mu$m pixels.  However at $f$1.5, the HERMES camera design is
slower than the $f$1.3 cameras of AAOmega, resulting in a broader PSF
(5\,pixel) than that of the currently implemented AAOmega modes.  This
results in an increased profile overlap compared to the basic AAOmega
mode.

\section{Fibre-to-Fibre cross contamination}
\label{need for crosstalk correction}
Before presenting our new extraction algorithm which reduces
fibre-to-fibre spectral cross-contamination (\S~\ref{the algoritham})
we will motivate its' development by consider the level of this {\it
cross-talk} in models of the four observing systems of
Table~\ref{Crosstalk modes}.  The pertinent parameters for the models
of each observing mode are:\\

\vspace{-0.25cm}\noindent{\it Fibre pitch - }The spacing between
fibres on the CCD.\\

\vspace{-0.25cm}\noindent{\it Fibre profile - }e.g.\ the projected
FWHM of a fibre profile on the CCD.\\

\vspace{-0.25cm}\noindent{\it Relative intensity - }The intensity
ratio between spectra to be extracted.

Fig.~\ref{Crosstalk profiles} presents a visual representation of the
instrument modes of Table~\ref{Crosstalk modes}.  A pair of isolated
fibres are considered.  This represents the limiting case of
cross-talk between two adjacent fibres only.  Two model fibre profiles
are shown with an input $\Delta$\,mag=3, a flux ratio of
ratio$\sim$15.8.  The fibre centre, and $\pm$3$\sigma$ range are
marked for each fibre (solid bars) as are the inter-fibre
ranges\footnote{We define the inter-fibre range for each fibre as
$\pm$half the fibre pitch, centred on the fibre.}  associated with
each fibre (dashed bars). This later range is used for the {\it
Tramline summation} extraction, while $\pm$3$\sigma$ is used for the
{\it Gaussian summation} unless it is larger than the inter-fibre
range, in which case it is truncated to this range.  A visual
inspection of Fig.~\ref{Crosstalk profiles} indicates that for all but
the default AAOmega mode there may be large fibre-to-fibre
cross-contamination, due to the significant overlap between fibre
profiles, if a simple pixel intensity summation extraction is used.
For the later instrument modes an extraction process is needed that
accounts for the flux of both fibres simultaneously and attempts to
correctly distribute the intensity information between the two fibre
profiles in the overlap region.  A quantitative analysis follows.

Examination of observational data from the AAOmega system indicates
that a Gaussian fibre profile is an acceptable first approximation.
We will use the Full Width at Half Maximum, FWHM{\footnote{The FWHM is
related to the Gaussian width, $\sigma$, as FWHM=$2\sqrt{2\ln{2}}$
$\times$ $\sigma$ $\sim$ $2.355\sigma$}, as the measure of profile
width in what follows.  We express relative spectral intensities in
terms of the magnitude difference of the integrated profile fluxes,
$\Delta$\,mag, for two astronomical objects illuminating the fibres.
Conventional wisdom for the AAOmega system has been to limit
$\Delta$\,mag$<$3.

Fig.~\ref{Crosstalk ratios} demonstrates that for the default
AAOmega-MOS mode of operation (the two top figures in each column)
cross-contamination of the fainter spectrum is limited to $<$1.6\% of
its integrated profile flux for $\Delta$mag$<$3.  Improved accuracy
requires a reduced $\Delta$\,mag.  However the three remaining
instrument modes detailed in \S~\ref{instrument modes} all suffer
significant cross-contamination.  In these cases not only is
cross-contamination of the weaker spectrum increased significantly,
but double counting of flux leads to $>$10\% errors in flux
normalisation when $\Delta$\,mag$<$0.3.  Clearly an improved
extraction methodology is needed which can account for the overlapping
fibre profiles in these cases.
No figure is given for the {\it Multi-profile deconvolution}
extraction since, under the assumption of this idealised initial
analysis, it returns perfect results for all four instrument modes.

\begin{table}
\caption{\label{Crosstalk modes} The pertinent parameters for the
four model instrument modes of the AAOmega system are given.}
\vspace{0.25cm}
\begin{center}
\begin{tabular}{|lccc|}
%{\bf Instrument} & {\bf Fibres} & {\bf Fibre pitch} & {\bf Fibre FWHM}\\
%{\bf Mode}       &              & {\bf (pixels)}    & {\bf (pixels)}\\ 
\hline
Instrument  & Fibres & Pitch & FWHM\\
Mode        &        &(pixels)    & (pixels)\\ 
\hline
AAOmega      & 392 & 10.0 & 3.4\\
Mini-Shuffle & 784 & 5.0  & 3.4\\
SPIRAL       & 512 & 4.0  & 2.4\\
\hline
HERMES       & 392 & 10.2 & 5\\ 
\hline
\end{tabular}
\end{center}
\end{table}

\clearpage 

\begin{figure}
\begin{center}
\includegraphics[width=7cm]{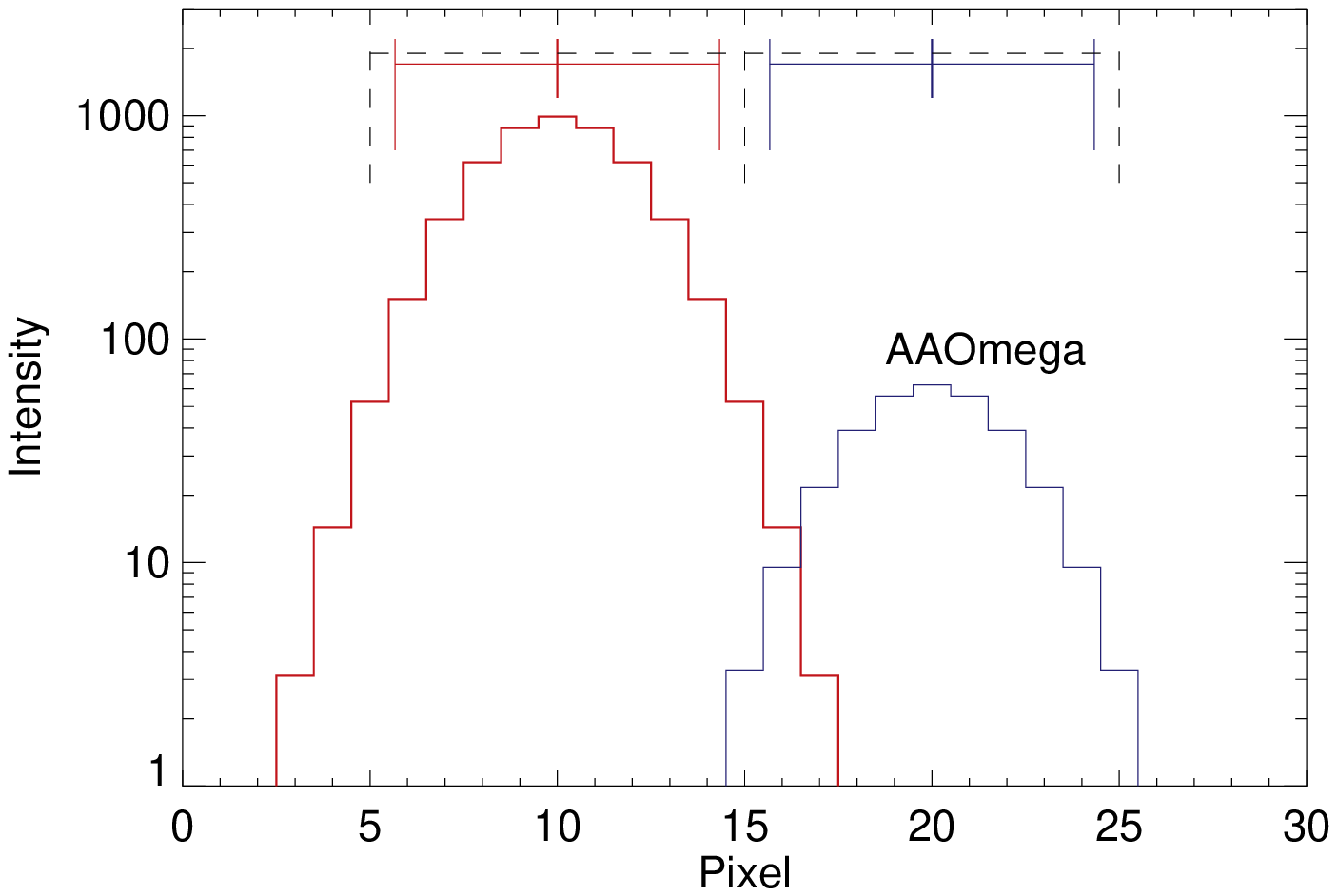}\\
\includegraphics[width=7cm]{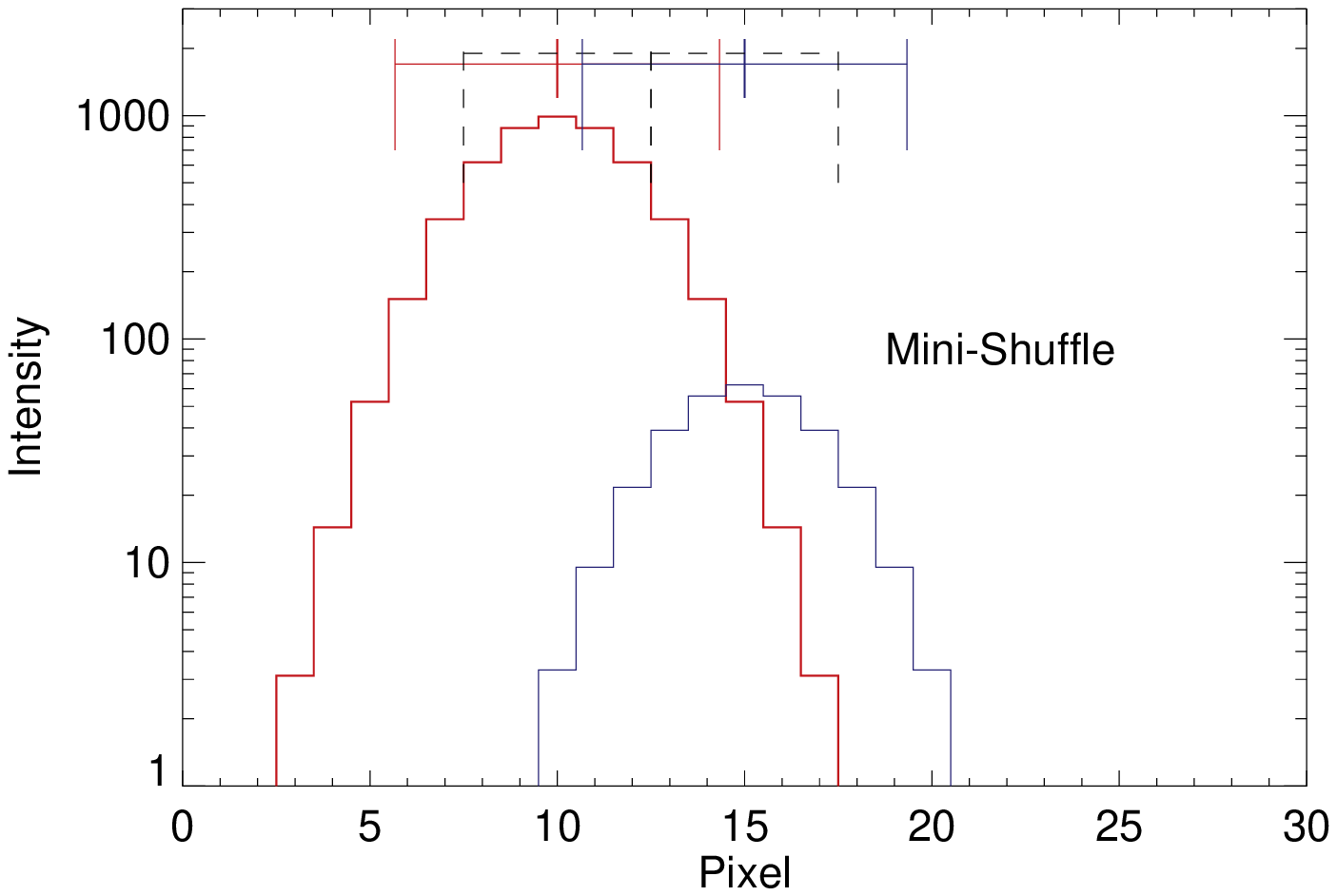}\\
\includegraphics[width=7cm]{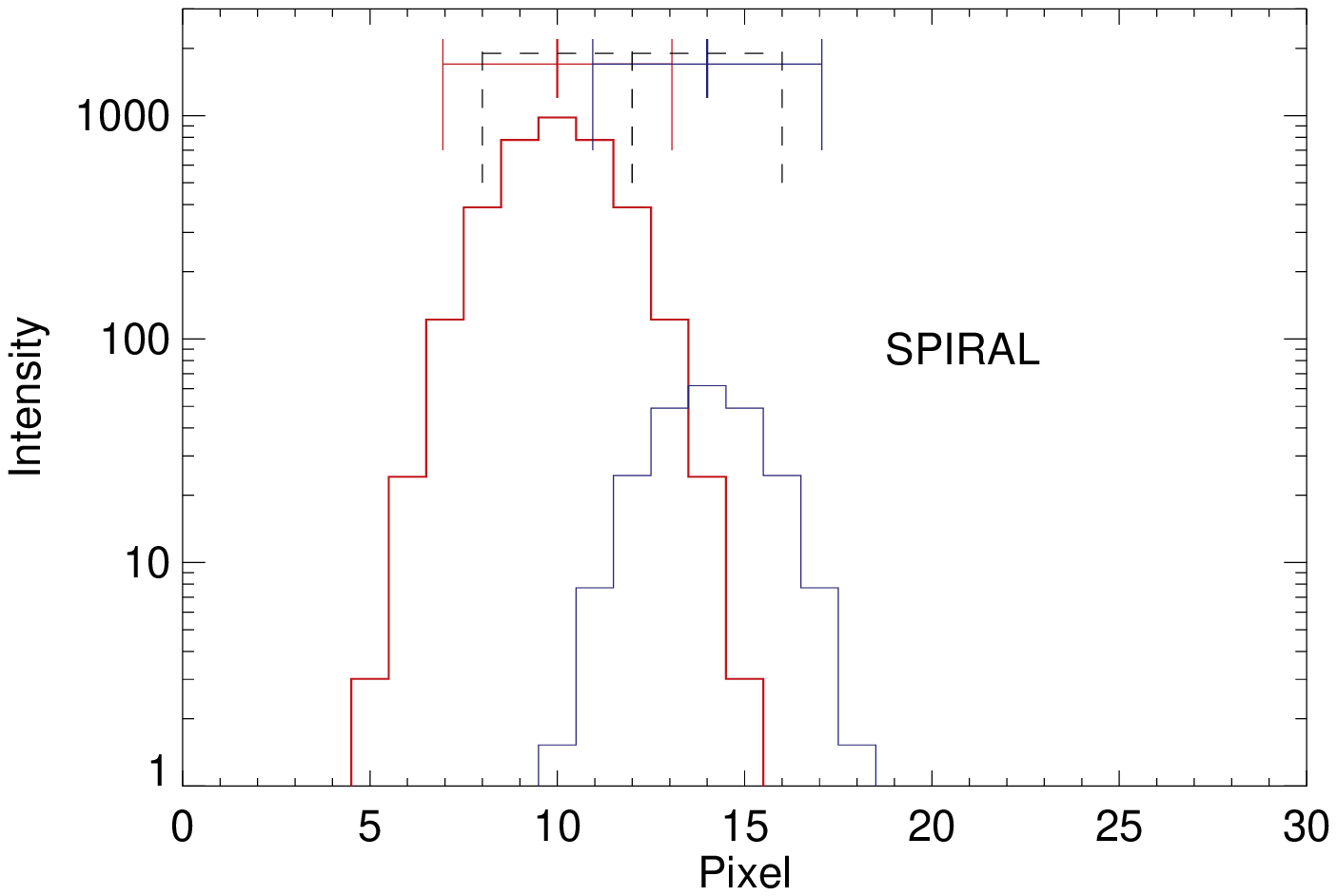}\\
\includegraphics[width=7cm]{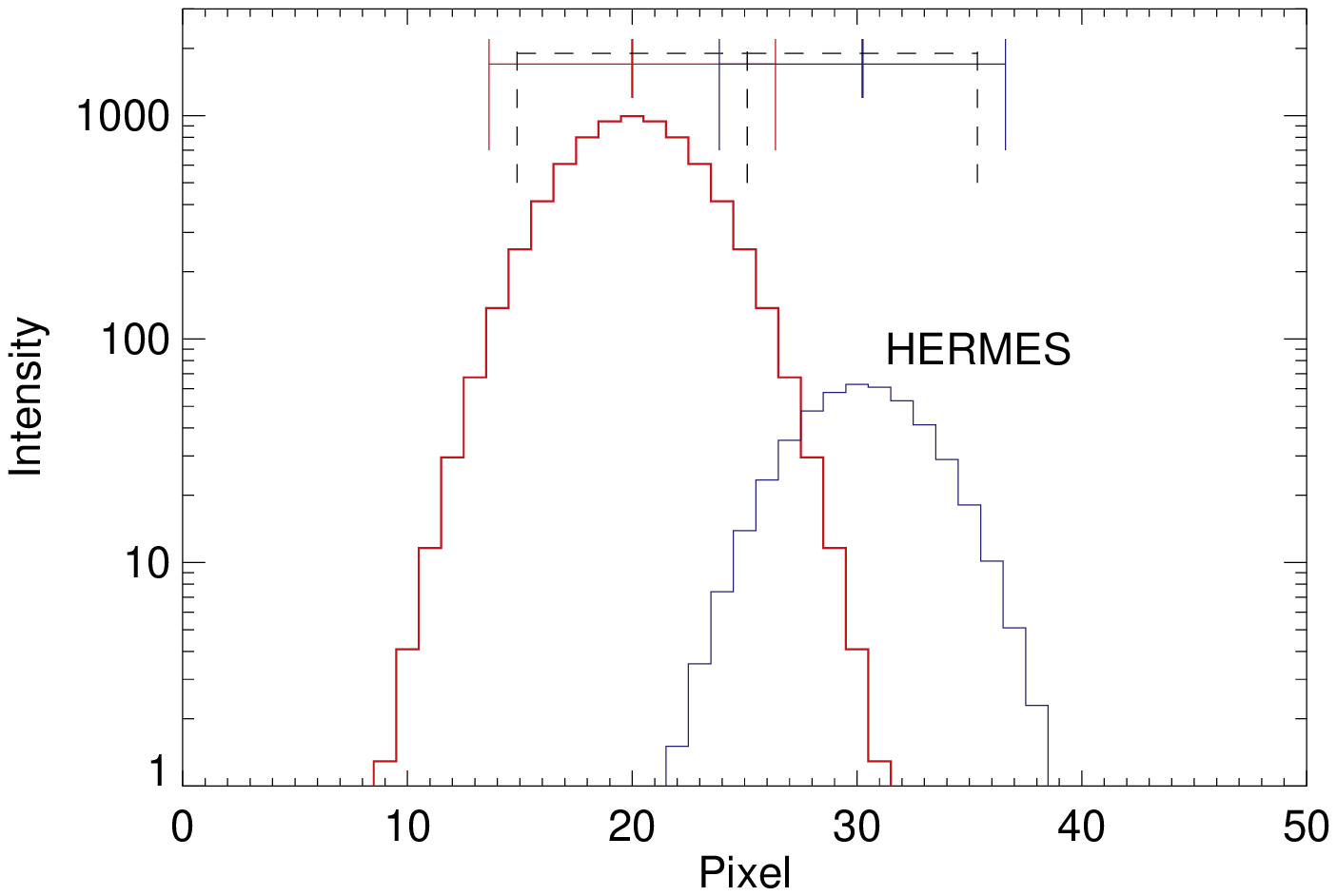}\\
\end{center}
\caption{\label{Crosstalk profiles} Model fibre profiles are shown for
a pair of fibres with each of the instrument modes given in
Table~\ref{Crosstalk modes}. The logarithmic scaling heightens the
visibility of the region of overlapping profiles. The profile
intensity ratio is $\sim$15.8 giving $\Delta$\,mag=3.  The fibre
centres, and $\pm$3$\sigma$ range are marked for each fibre (solid
bars) as are the free inter-fibre ranges associated with each fibre
(dashed bars).  For all but the first instrument profile, the
3$\sigma$ range extends beyond the inter-fibre range.}
\end{figure}

\clearpage

\begin{figure*}
\begin{center}
\includegraphics[width=7cm]{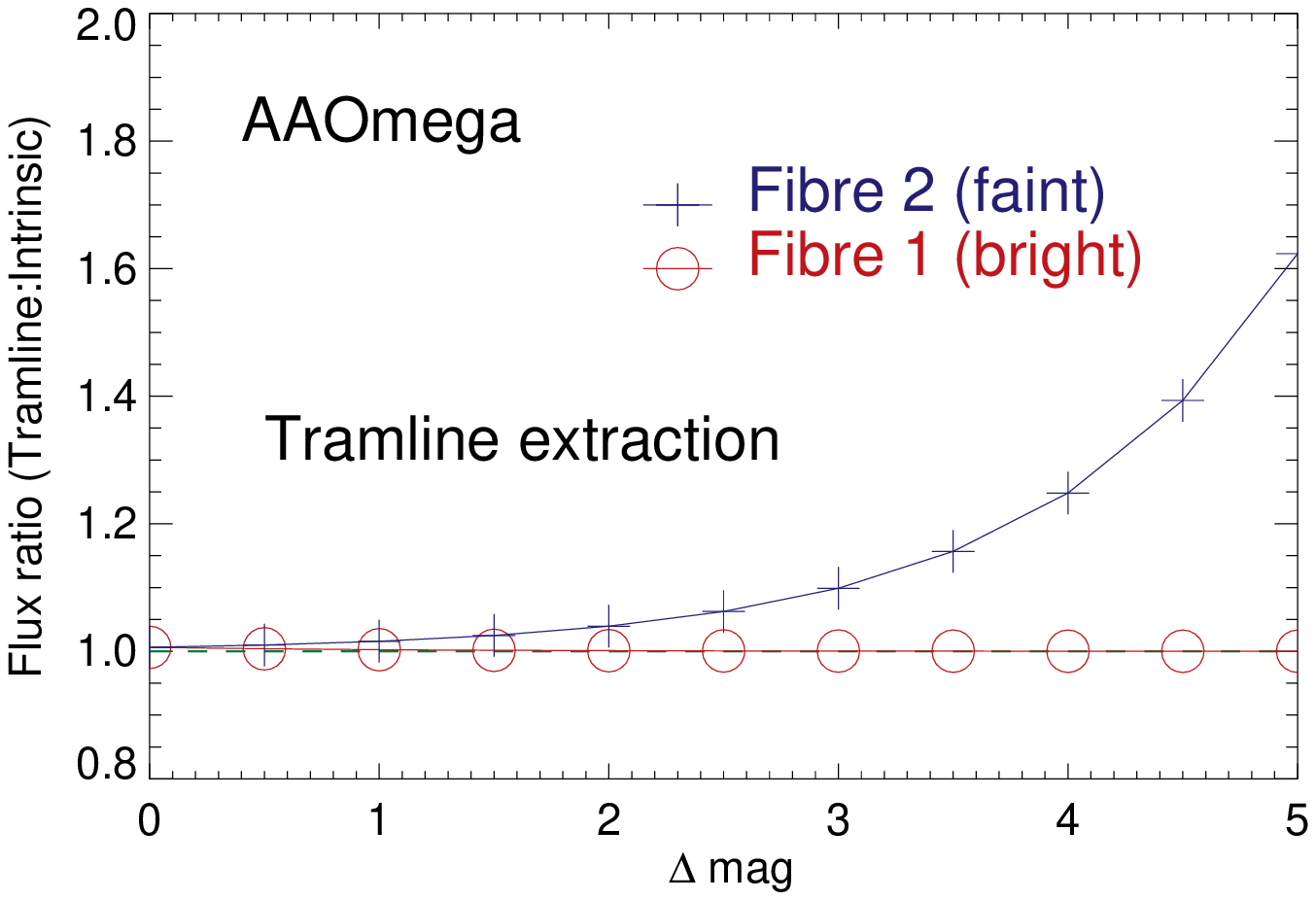}
\includegraphics[width=7cm]{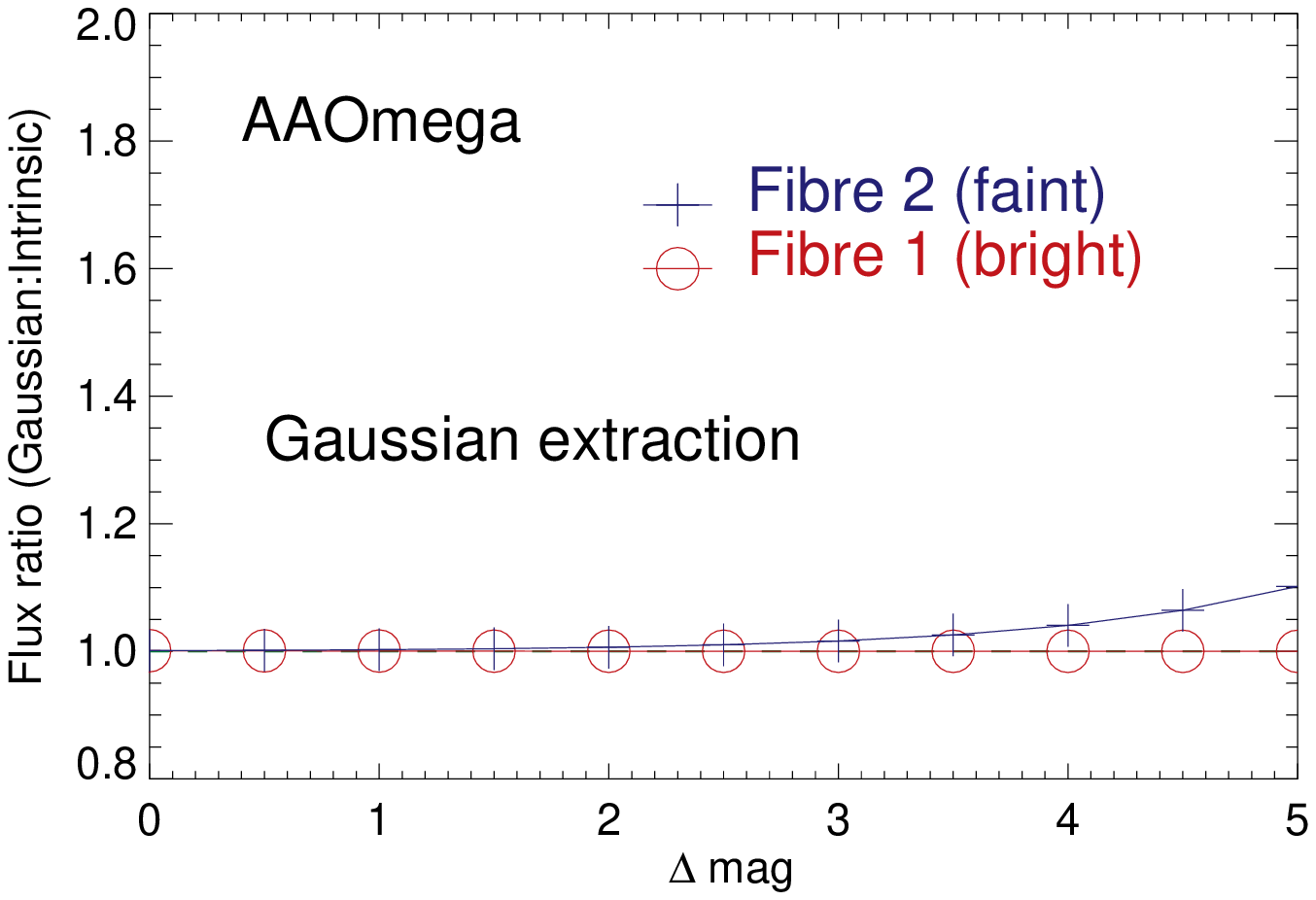}\\
\includegraphics[width=7cm]{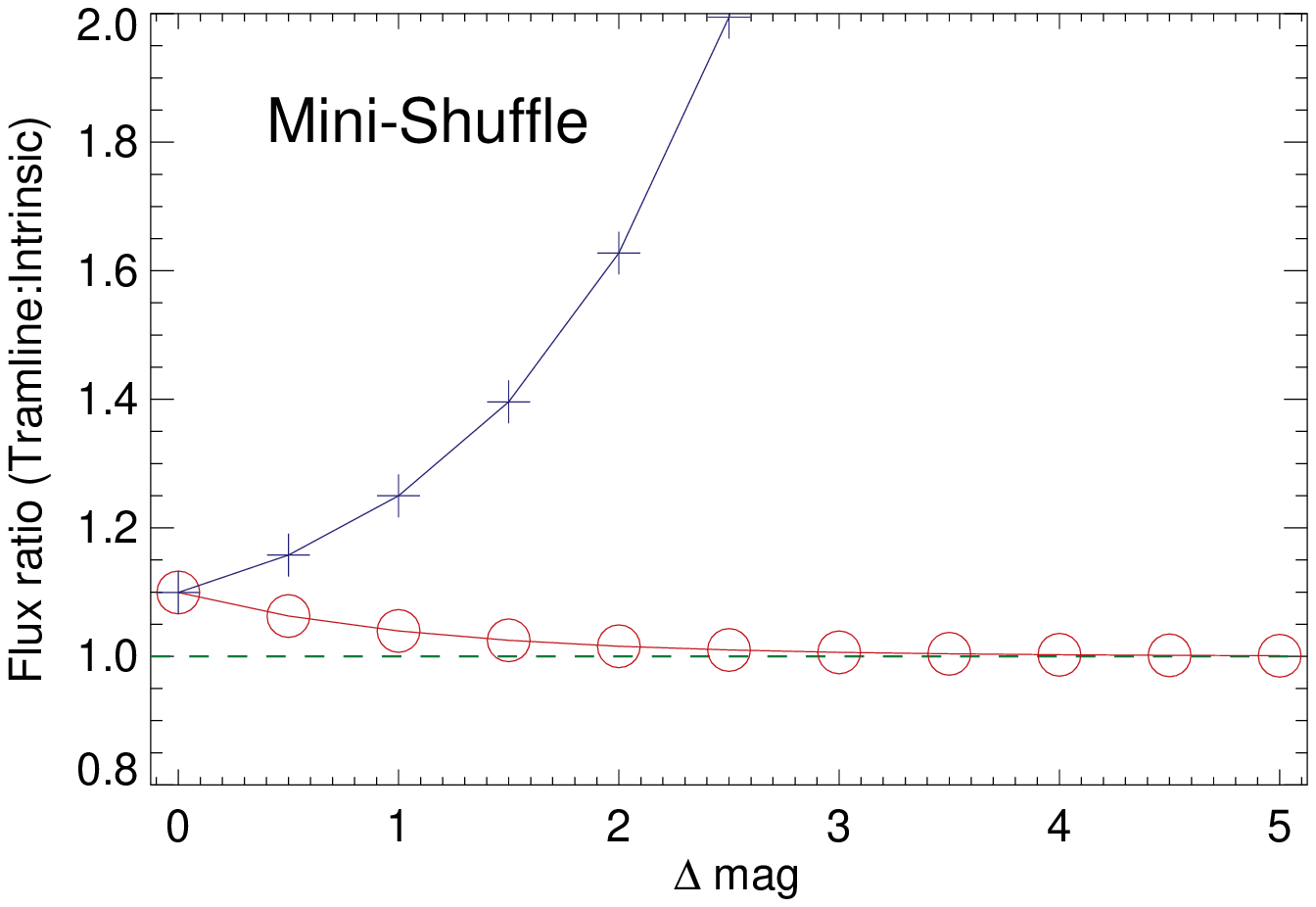}
\includegraphics[width=7cm]{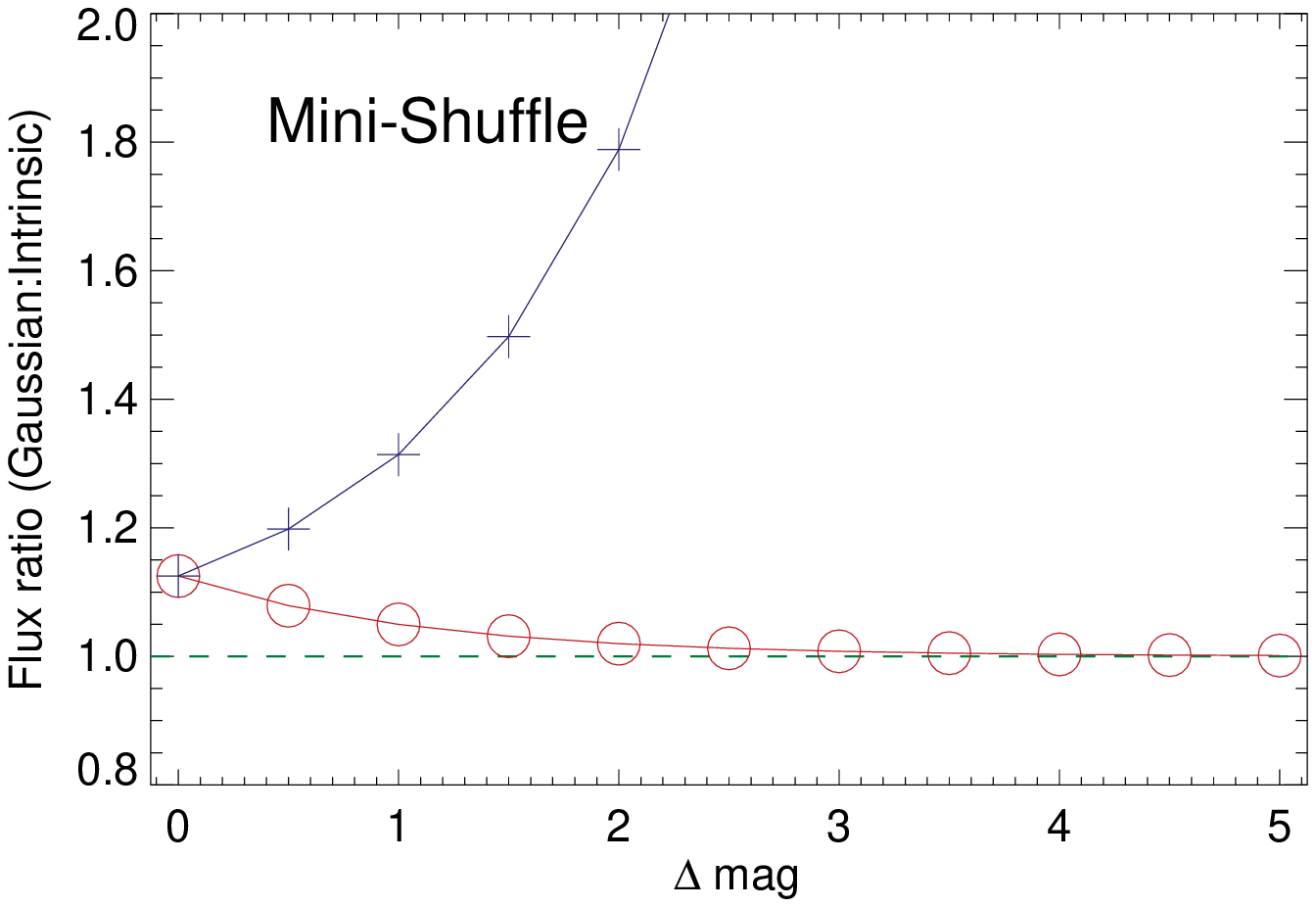}\\
\includegraphics[width=7cm]{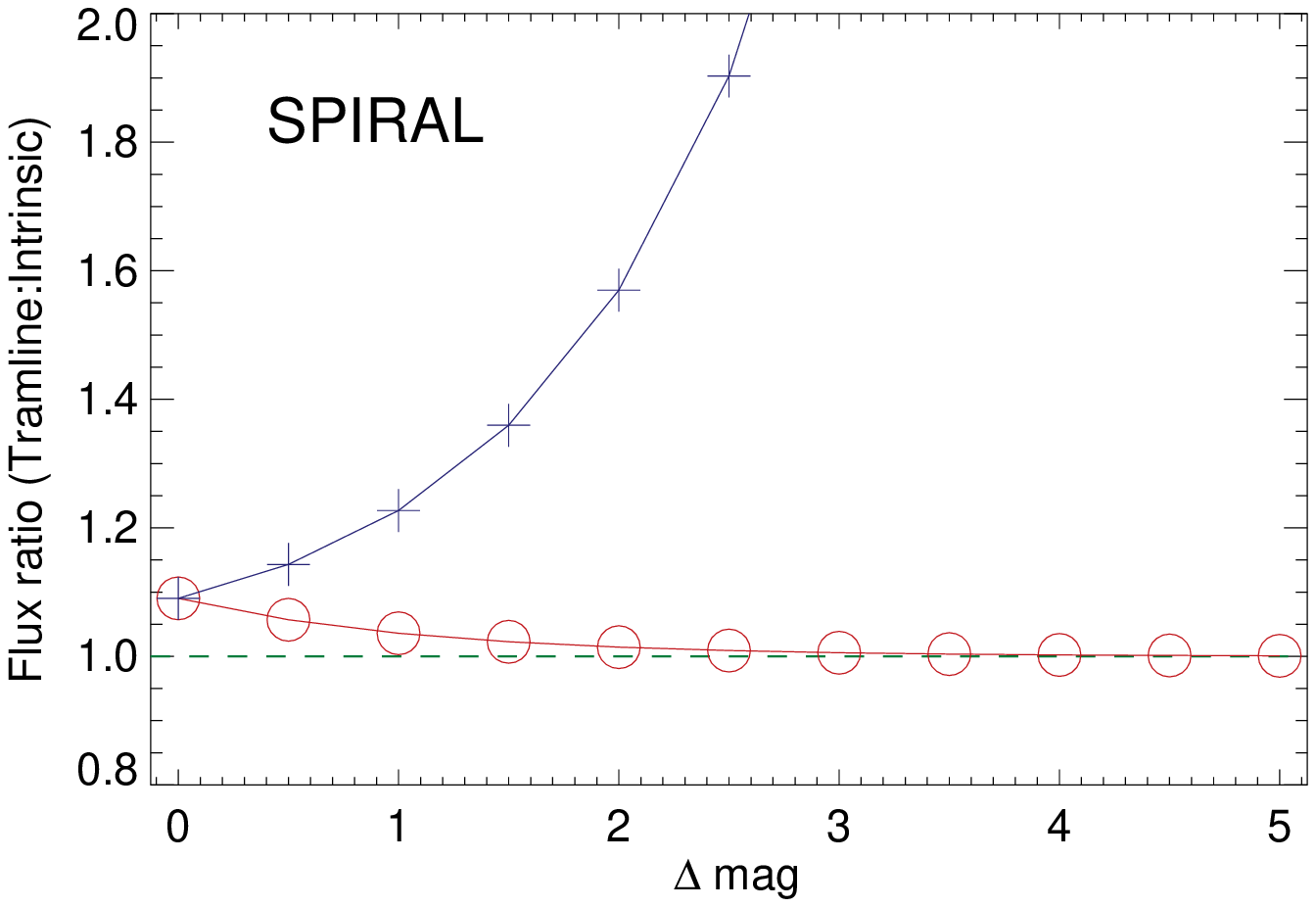}
\includegraphics[width=7cm]{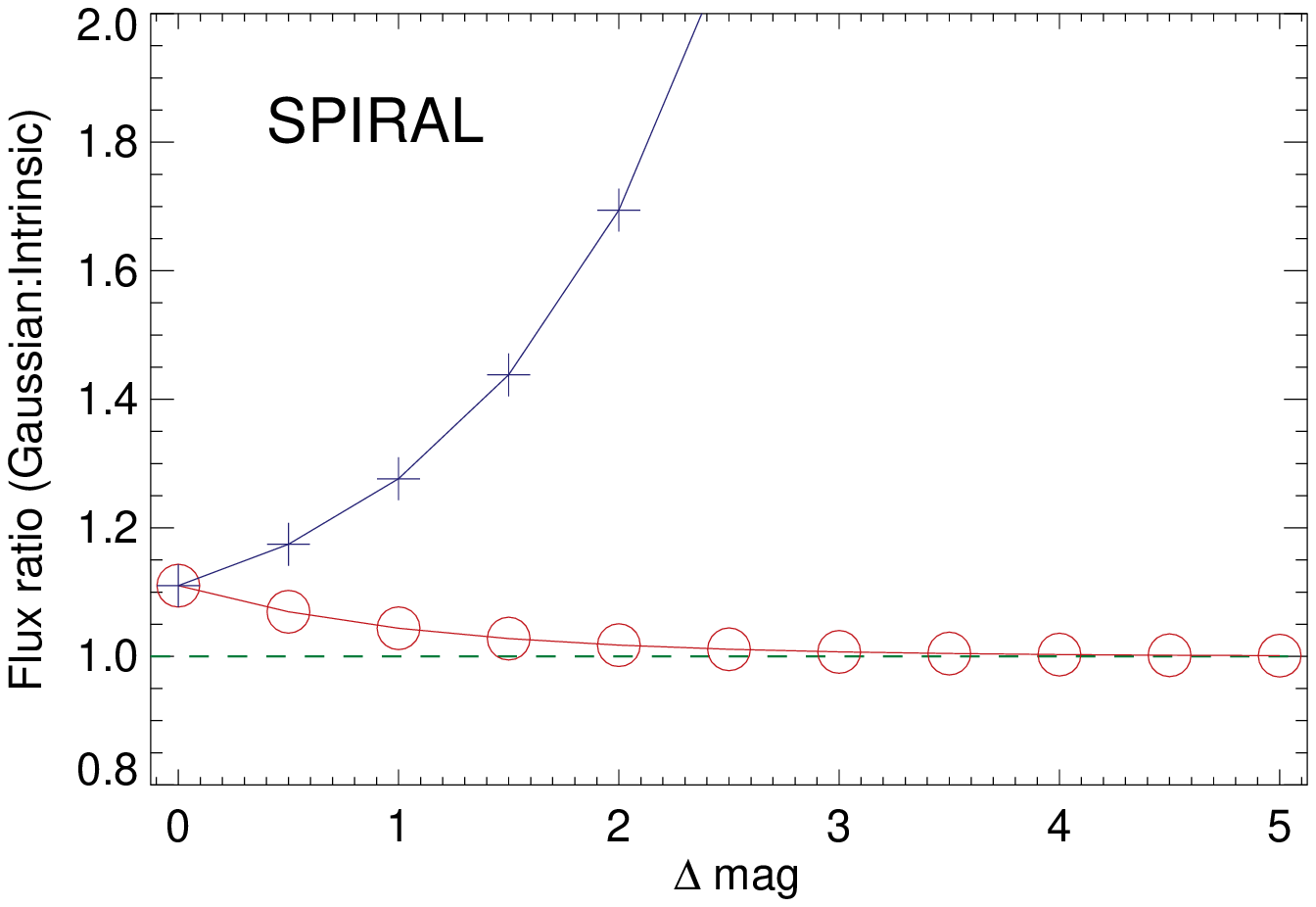}\\
\includegraphics[width=7cm]{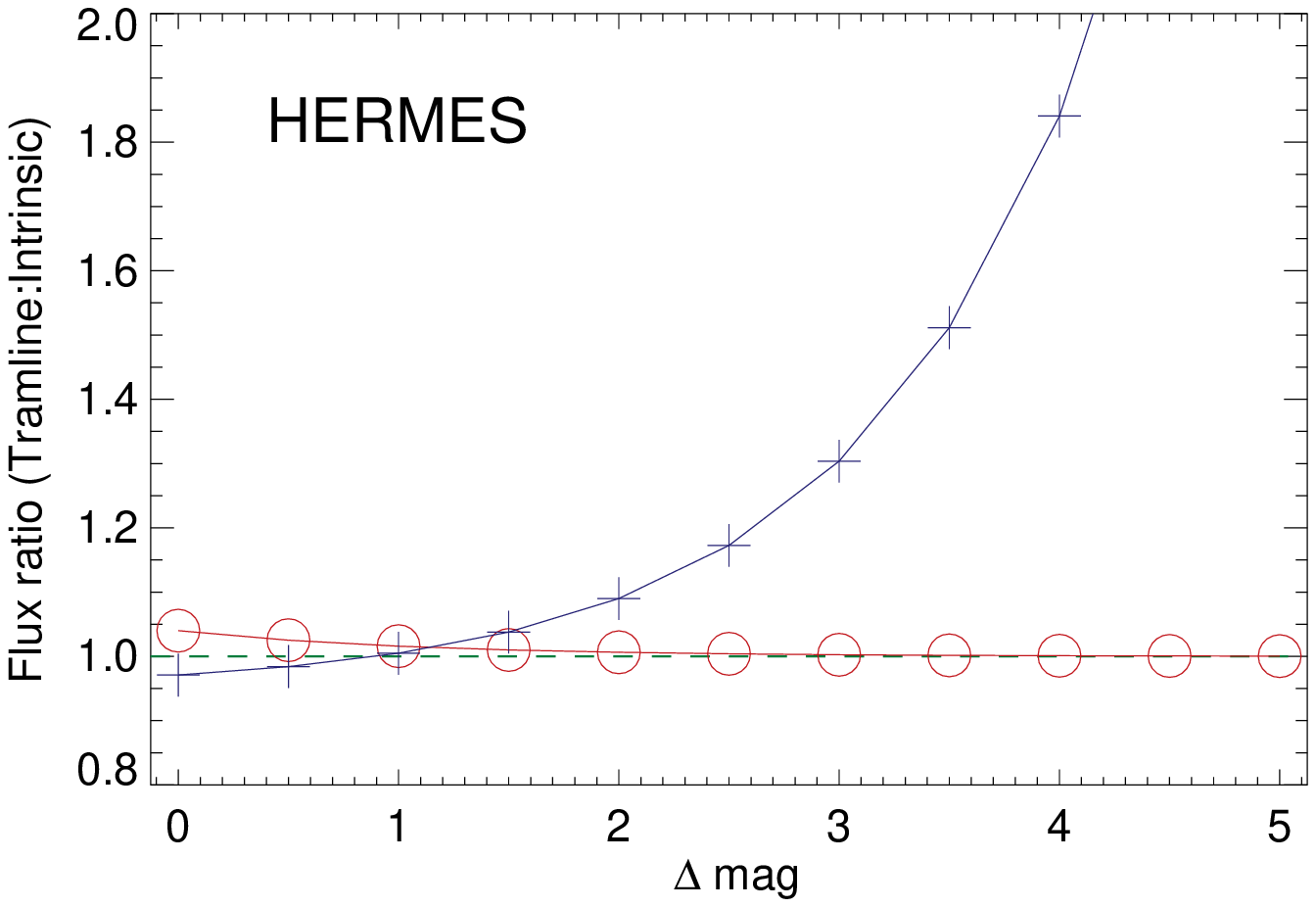}
\includegraphics[width=7cm]{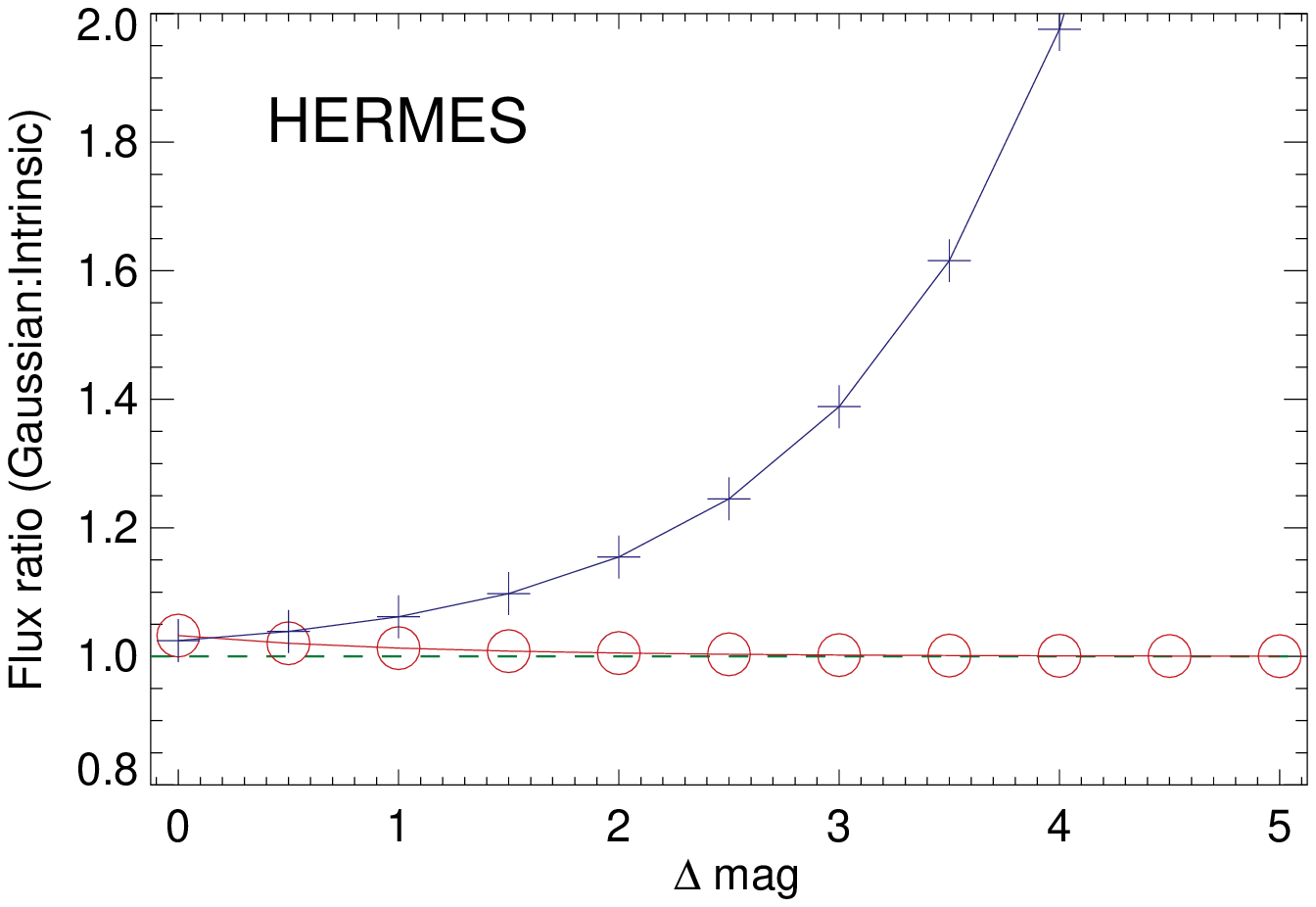}\\
\end{center}
\caption{\label{Crosstalk ratios} Using two model Gaussian fibre
profiles and the instrument modes of Table~\ref{Crosstalk profiles}
the ratio of extracted flux to input model profile flux is presented
as a function of the relative intensity difference (expressed as a
magnitude difference, $\Delta$\,mag) between the two input profiles.
The left column shows results using the Tramline summation extraction,
the right the Gaussian least squares fit.  In all cases the second
fibre is $\Delta$\,mag fainter than the brighter first fibre.  An
accurate extraction would follow the 1:1 locus.  For all instrument
modes the extracted flux for fibre 2, the fainter profile, is
increasingly in error as $\Delta$\,mag increases due to
cross-contamination from the brighter source fibre.
\smallskip
No traces are shown for the {\it multi-profile deconvolution}
extraction as the process is found to be 100\% accurate (the correct
flux is recovered for both fibres) in the noise free model limit for
all four instrument modes.}
\end{figure*}
\clearpage

\subsection{Effects of errors in centroids and profile widths}
Two parameters of the assumed fibre profiles dominate the extraction
error, errors in the assumed fibre centroid and errors in the profile
widths.  Fig.~\ref{Errors} demonstrates the effects of introducing
random modifications to the fibre centroids and widths (in isolation).
Fifty thousand realisations are computed with a random error
introduced to each fibre profile drawn from a normal distribution with
the indicated width.  The ratio of the mean extracted flux to the
input model value (and associated $\pm$1-$\sigma$ scatter) is given
for each error distribution.  The flux ratio between the fibres was
set to unity ($\Delta$\,mag=0).  The simulations show that while
errors in the fibre centroid value may be tolerated at the 0.2\,pixel
level (6\% of the FWHM here), an accurate knowledge ($<$0.1\,pixel,
$\sim$3\% of FWHM) of the fibre profile is required to control errors
under all extraction methodologies.

\begin{figure}
\begin{center}
\includegraphics[width=7cm]{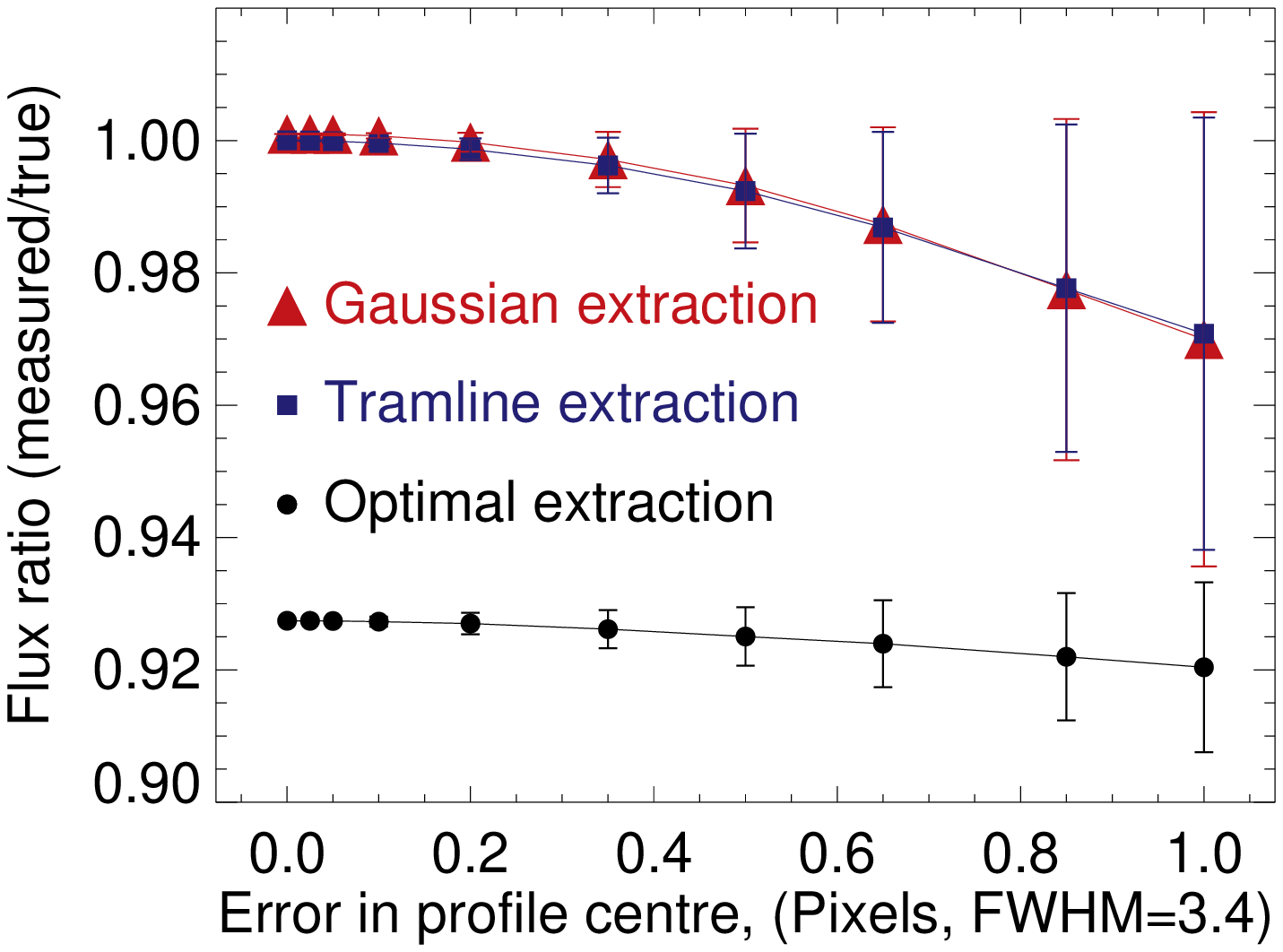}\\
\includegraphics[width=7cm]{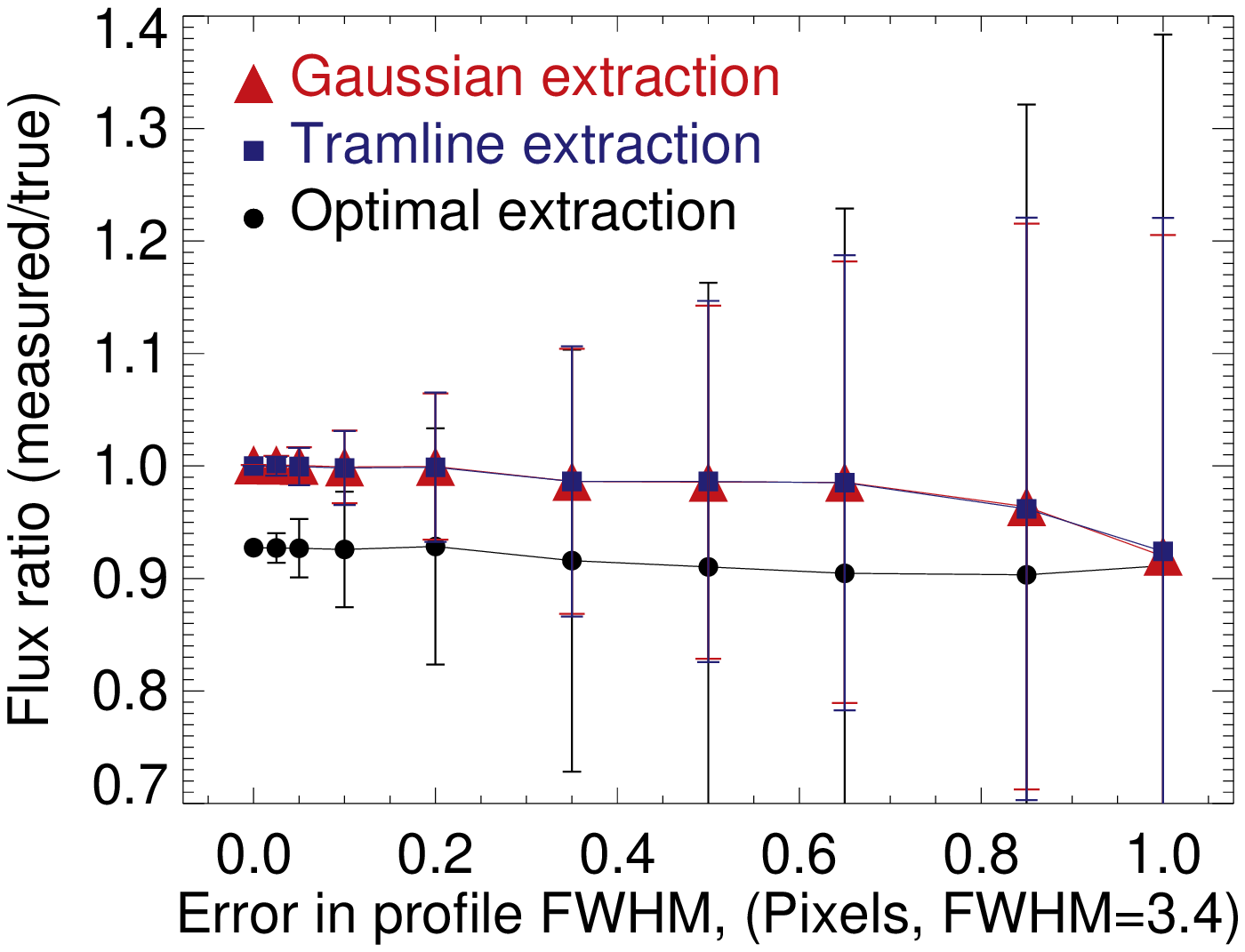}
\end{center}
\caption{\label{Errors} The ratio of the mean measured flux to the
input model intensity (and associated RMS scatter) is shown for the
three extraction methods of \S~\ref{spec ext} after 5,000 independent
realisations of the error model. For each realisation the model fibre
profile centroid and FWHM parameters are drawn from normal
distributions with the indicated widths.  The summation aperture
correction factor has not been applied to the {\it Tramline}
extraction, leading to its' offset below the other two extraction
methods.}
\end{figure}

\subsection{Effect of Signal-to-Noise ratio}
A final parameter to consider when testing the accuracy of the
extraction of fibre profiles is the signal-to-noise ratio (SN) of the
data.  To investigate this, he two fibre profile model is retained,
but a constant readout noise per pixel and a shot-noise component is
added to each realisation of the model data based on the intrinsic
strengths of the input model profiles.  Profiles are generated with a
range of normalisations while the input magnitude ratio
($\Delta$\,mag) is held fixed for each realisation.  Fig.~\ref{SN
Errors AAOmega} shows the results obtained for the three extraction
methodologies, for a range of SN ratios, assuming the default
AAOmega-MOS mode as described in Table~\ref{Crosstalk modes}.

The figure shows the intrinsic signal-to-noise ratio used when
generating the model profiles and the signal-to-noise ratio of the
extracted profile flux estimate.  This recovered signal-to-noise is
defined as the ratio of the extracted count rate to the noise level as
estimated from the model data by the extraction process.  When
reviewing the results of the simulations, one should note that the SN
ratio recovered for each spectrum is a statistical measure derived
from the data, and does not directly address the issue of the double
counting of flux due to fibre-to-fibre crosstalk.

Fig.~\ref{SN Errors AAOmega} indicates that for the default
AAOmega-MOS mode the choice of extraction methodology has only limited
effect on the final SN ratio.  In the case of $\Delta\,$mag=0 there is
some indication that the multi-component decomposition (optimal
extraction) reduces the scatter in the ratio between recovered and
intrinsic profile flux.  This increase becomes more pronounced for
lower SN ratio data.  The situation is similar for $\Delta\,$mag=3
with the error in the flux value derived for the fainter fibre
suffering a greater inflation for the tramline and Gaussian extraction
in comparison to the optimal extraction.

The process is repeated for each of the instrument configurations from
Table~\ref{Crosstalk modes}.  Fig.~\ref{SN Errors all} shows the
results for $\Delta$\,mag=0 and Fig.~\ref{SN Errors all D3} for
$\Delta$\,mag=3.  It is immediately apparent that in the case of
tightly packed fibres fibre-to-fibre cross-talk plays a significant
role in modifying the extracted flux values for the tramline and
Gaussian extraction methods.

\begin{figure*}
\begin{center}
\includegraphics[width=7.5cm]{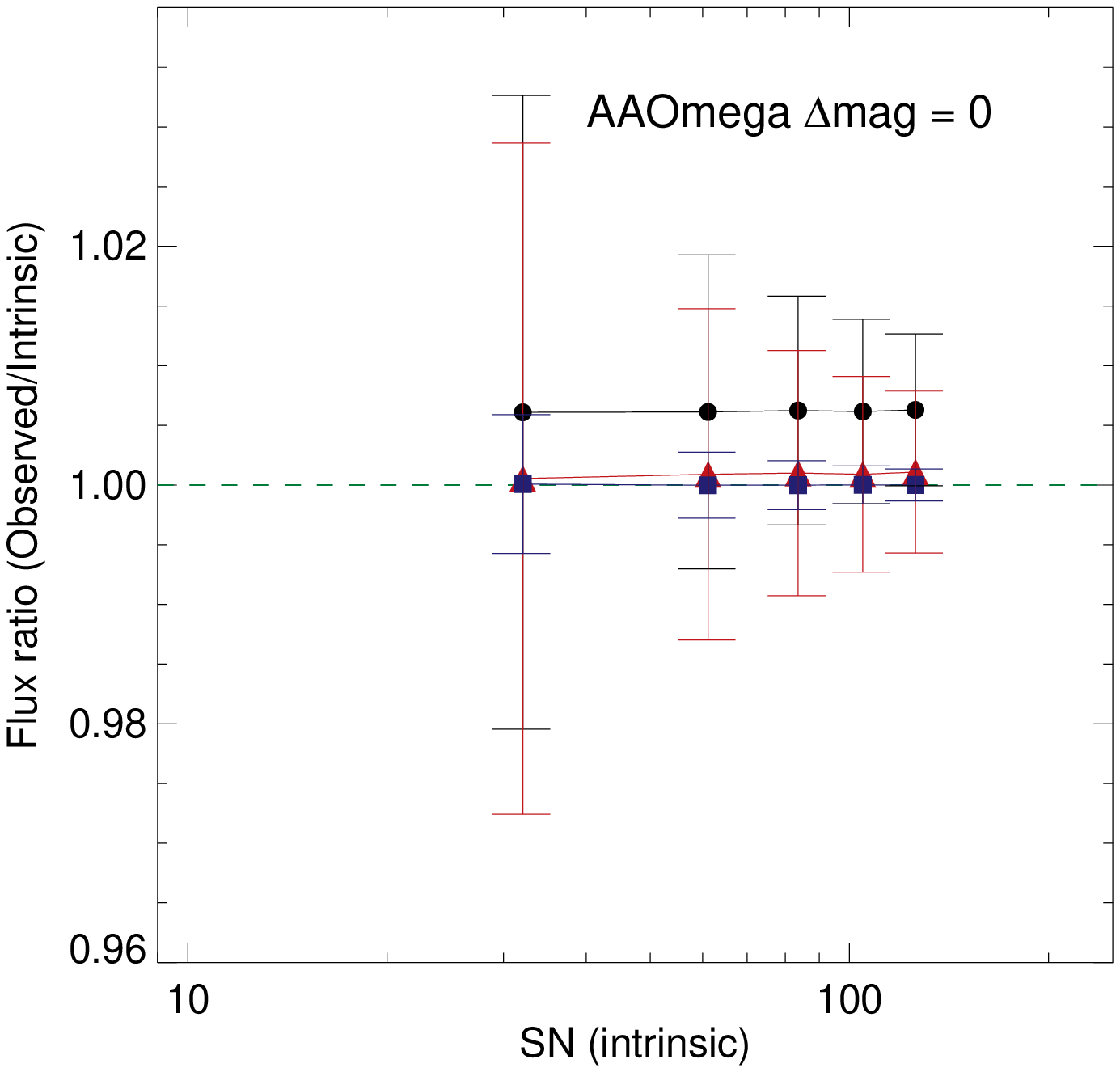}
\includegraphics[width=7.5cm]{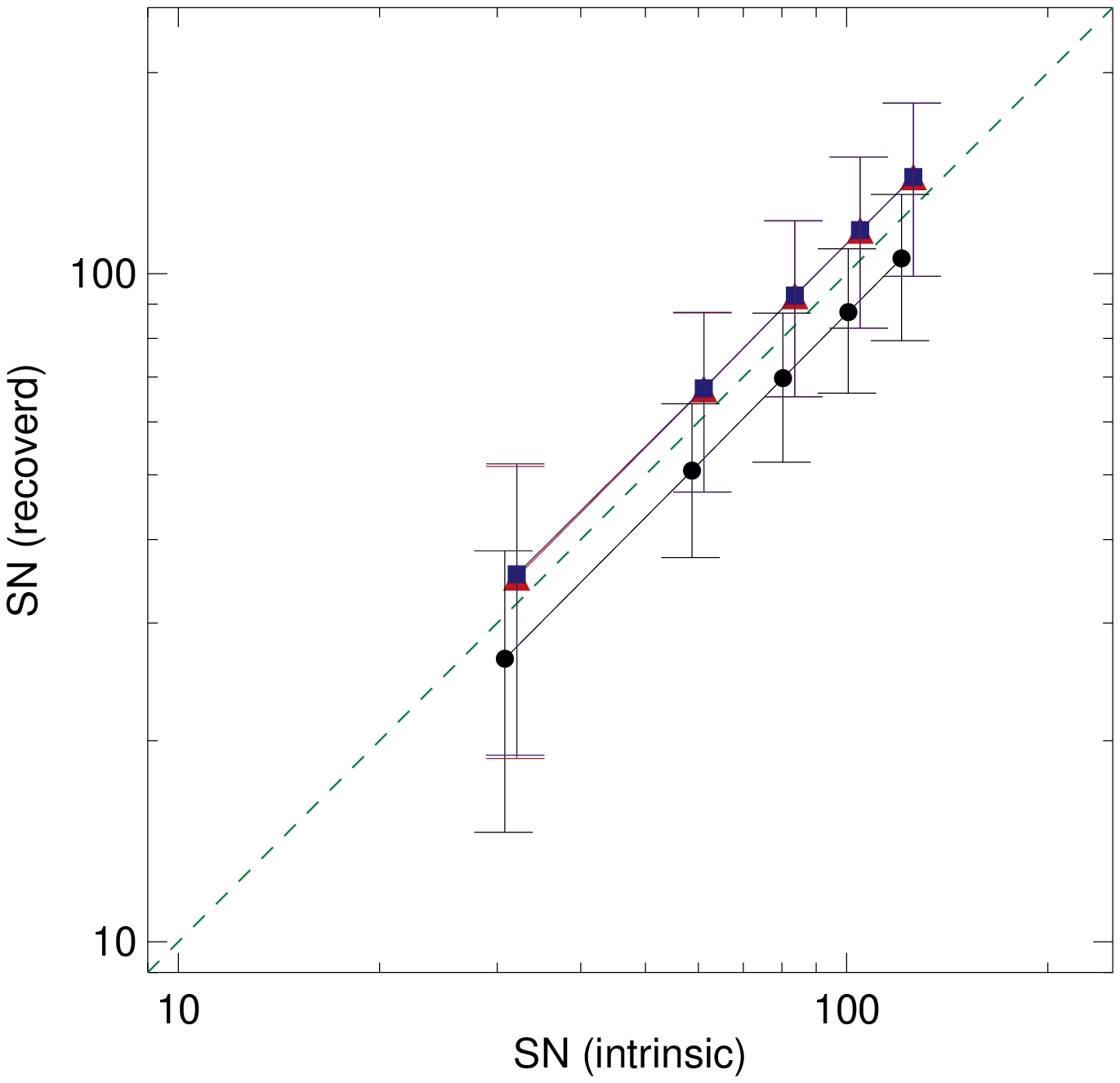}\\
\includegraphics[width=7.5cm]{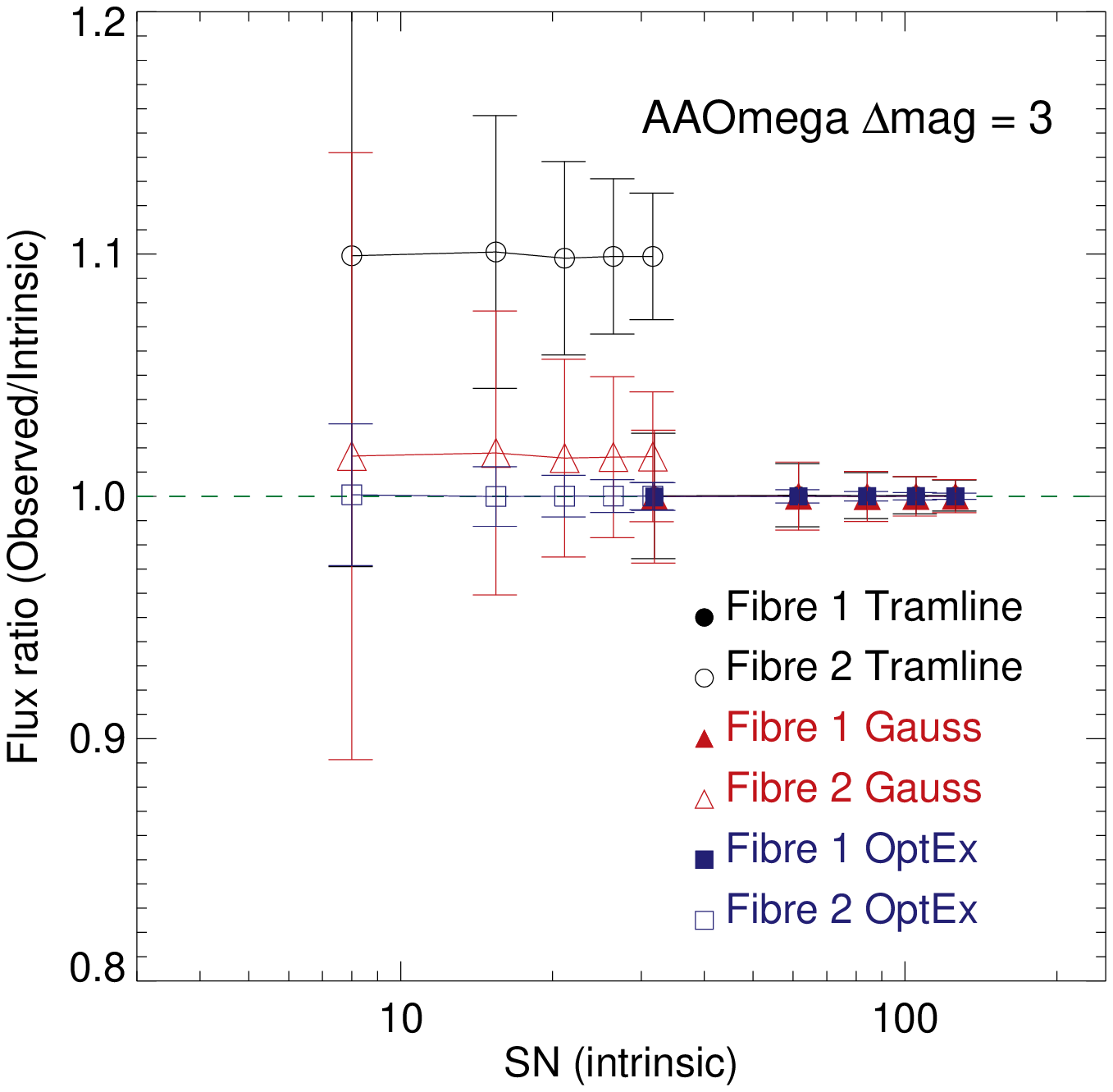}
\includegraphics[width=7.5cm]{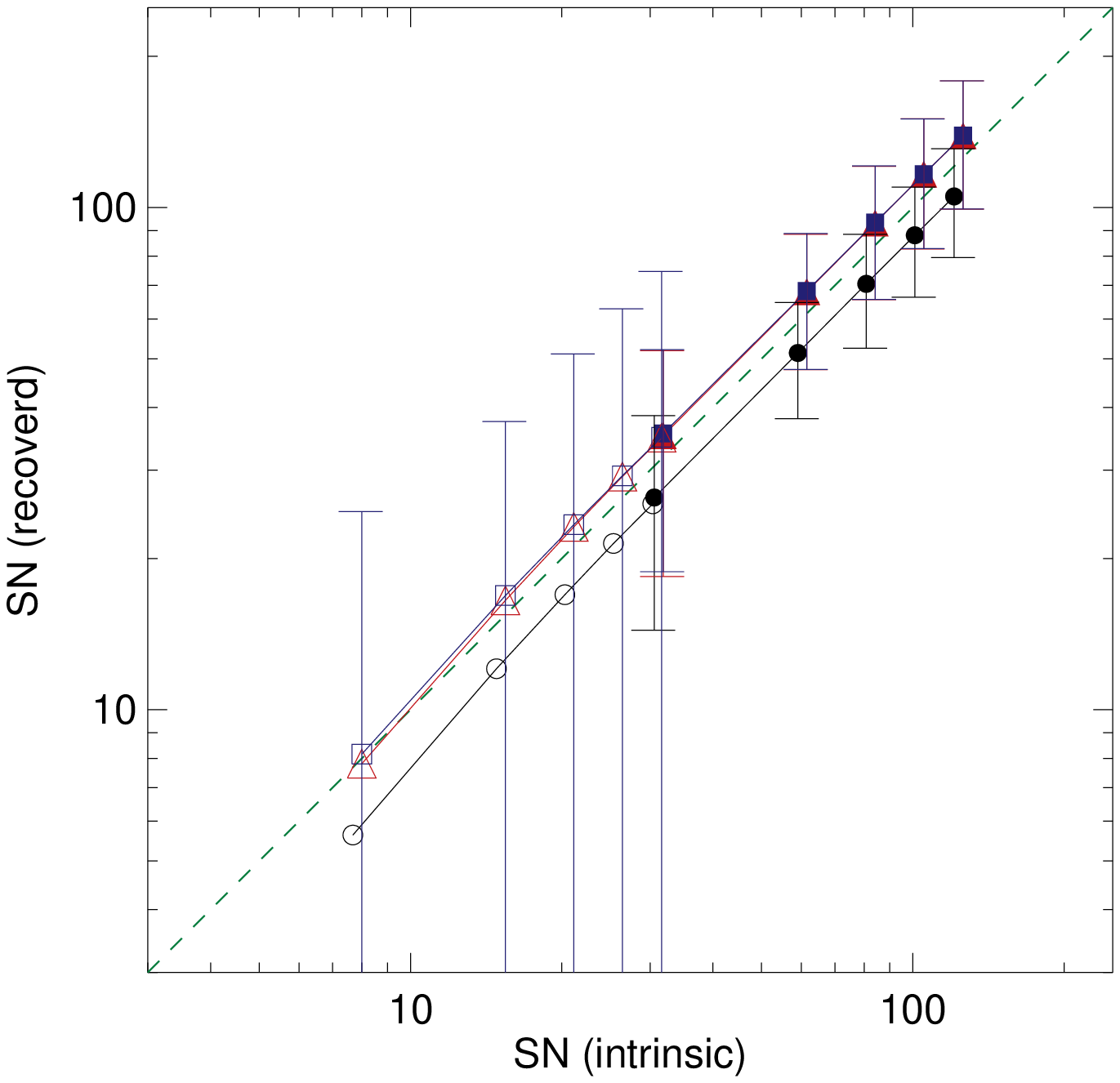}\\
\end{center}
\caption{\label{SN Errors AAOmega} Variations in the extracted flux
are considered as a function of the signal-to-noise ratio intrinsic to
the model date realisation.  Signal-to-noise is quoted for the
integrated profile flux.  Two sets of simulations are shown, both for
the AAOmega-MOS instrument mode.  Fifty thousand independent trials of
the noise model are run with the fibre profile flux ratio set to
$\Delta$\,mag=0 \& 3 (top and bottom rows respectively).  Fibre 2 is
the fainter fibre in the $\Delta$\,mag=3 case.  The left column
presents the ratio of the extracted flux to that intrinsic to the
model.  The right column compares the intrinsic SN to that recovered
for the extracted profiles.  In the case of $\Delta$\,mag=0 the
results for the two fibres are indistinguishable. The optimal
extraction is seen to return a smaller scatter in the recovered flux
ratio, the scatter remaining largely constant as SN decreases.  For
the $\Delta$\,mag=3 case both the Tramline and Gaussian extraction
return an increased scatter in the ratio for the fainter fibre (fibre
2) of the pair.  The RMS error bars for fibre two in the lower right
plot are similar for all three extraction algorithms and so for
clarity only those associated with the OptEx algorithm are given.}
\end{figure*}

\begin{figure*}
\begin{center}
\includegraphics[width=7.5cm]{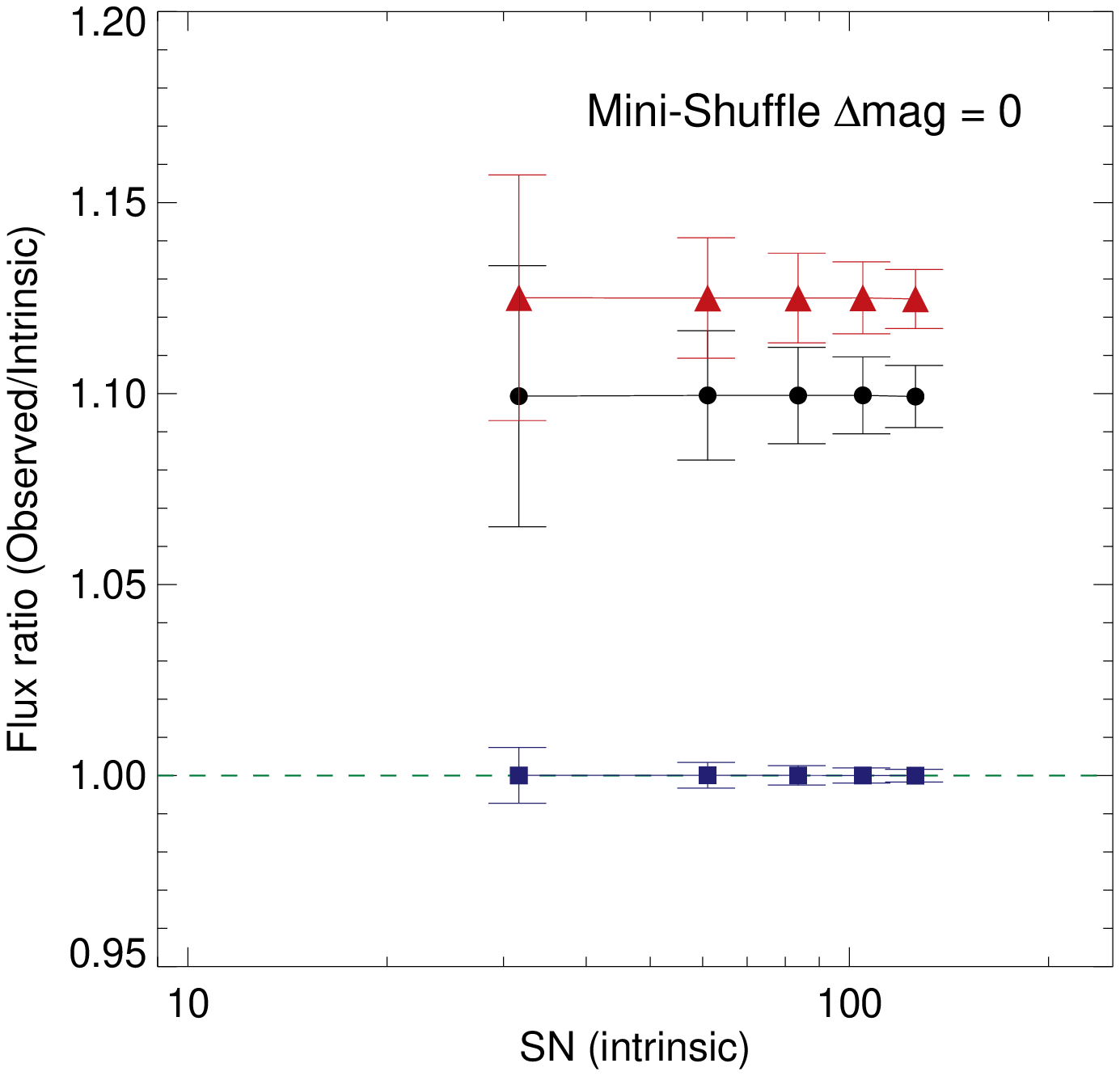}
\includegraphics[width=7.5cm]{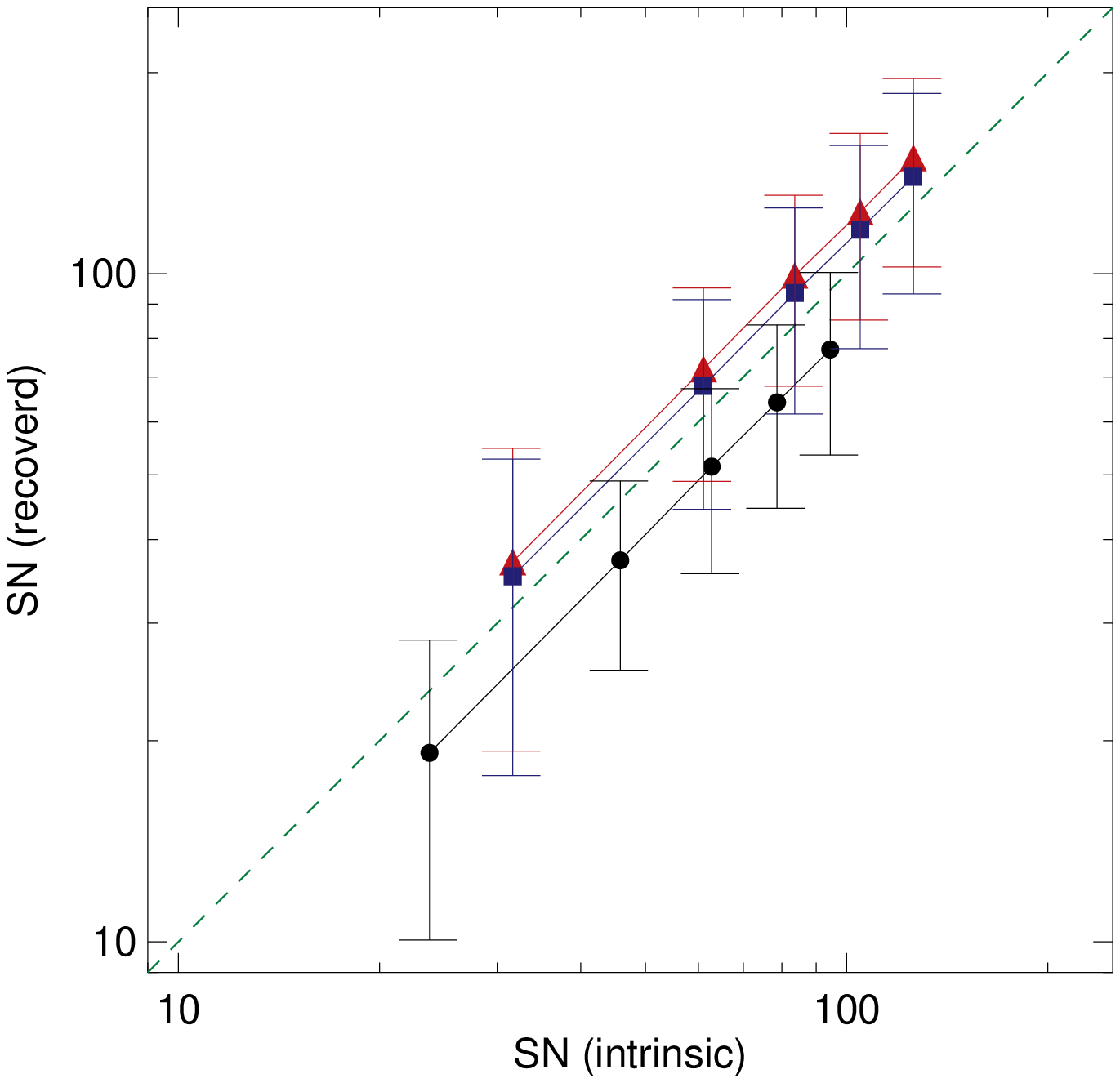}\\
\includegraphics[width=7.5cm]{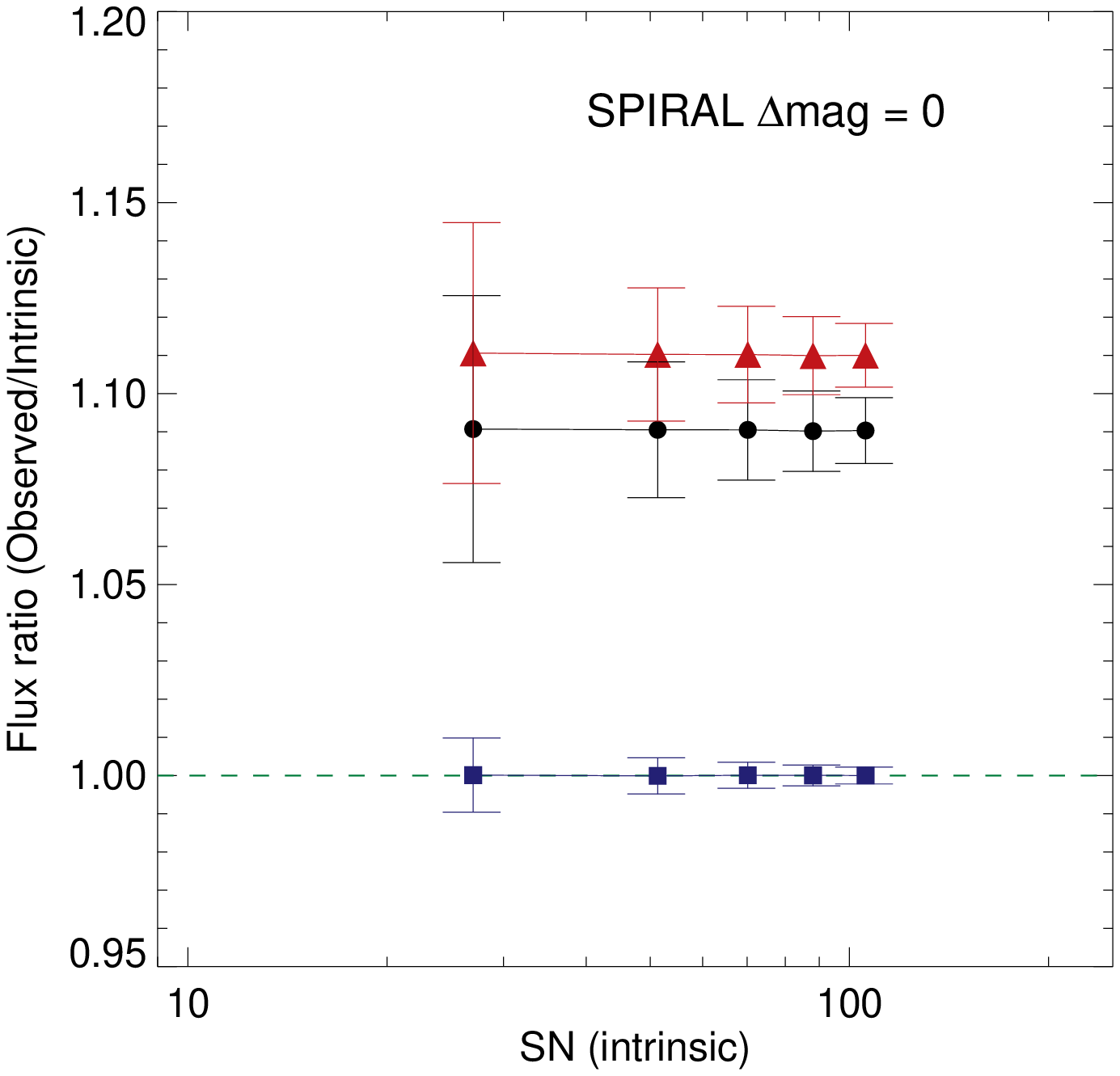}
\includegraphics[width=7.5cm]{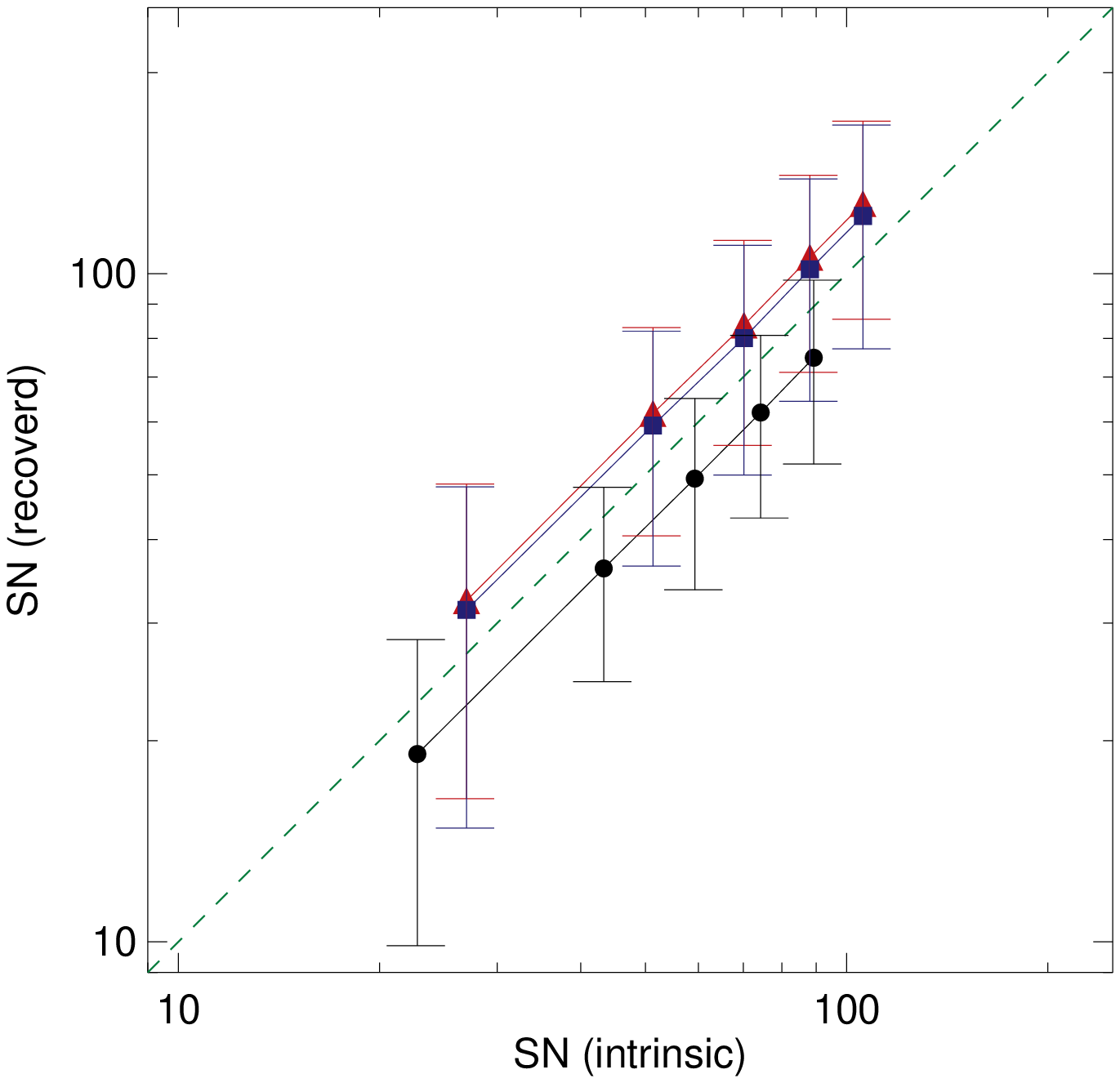}\\
\includegraphics[width=7.5cm]{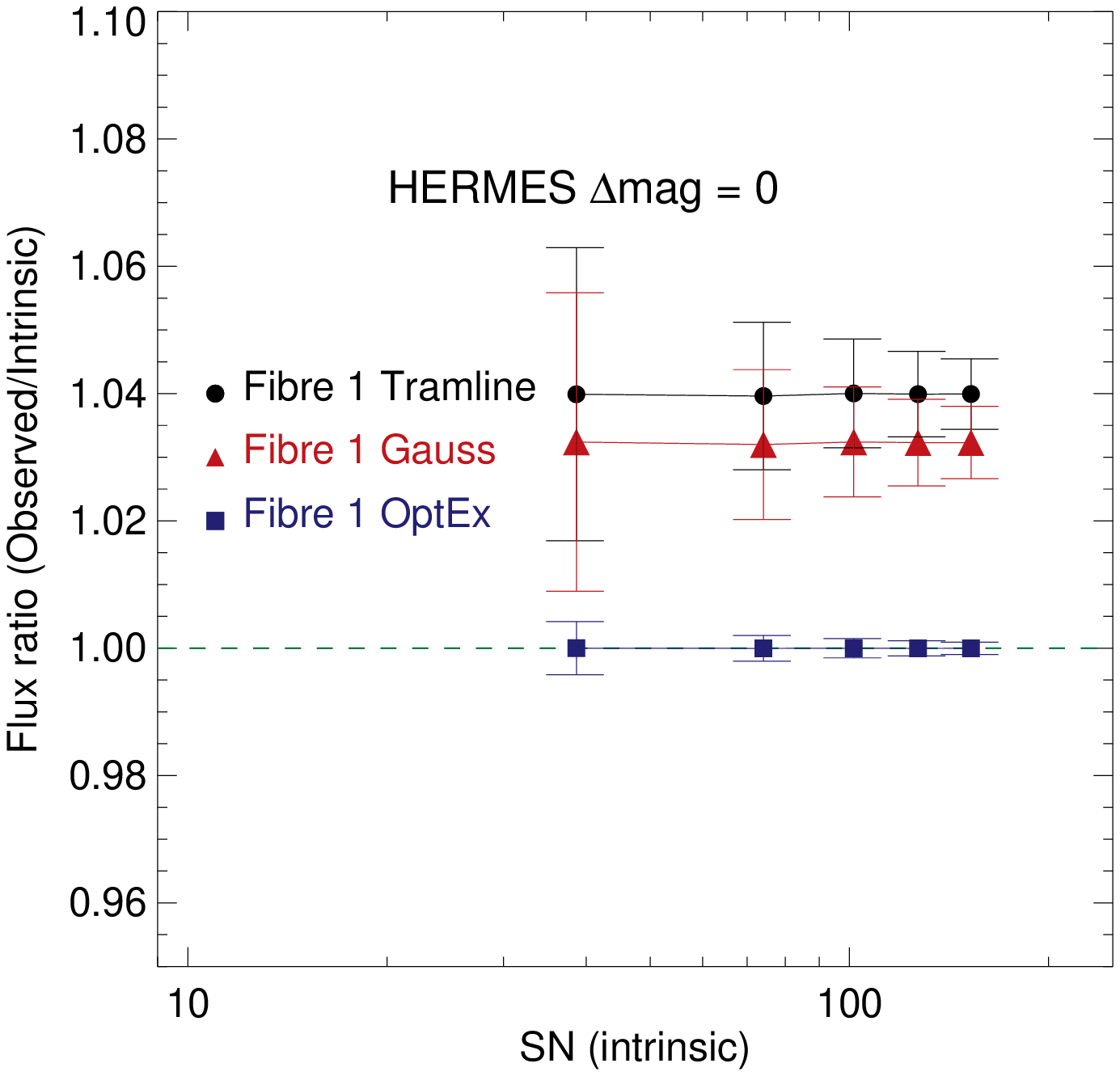}
\includegraphics[width=7.5cm]{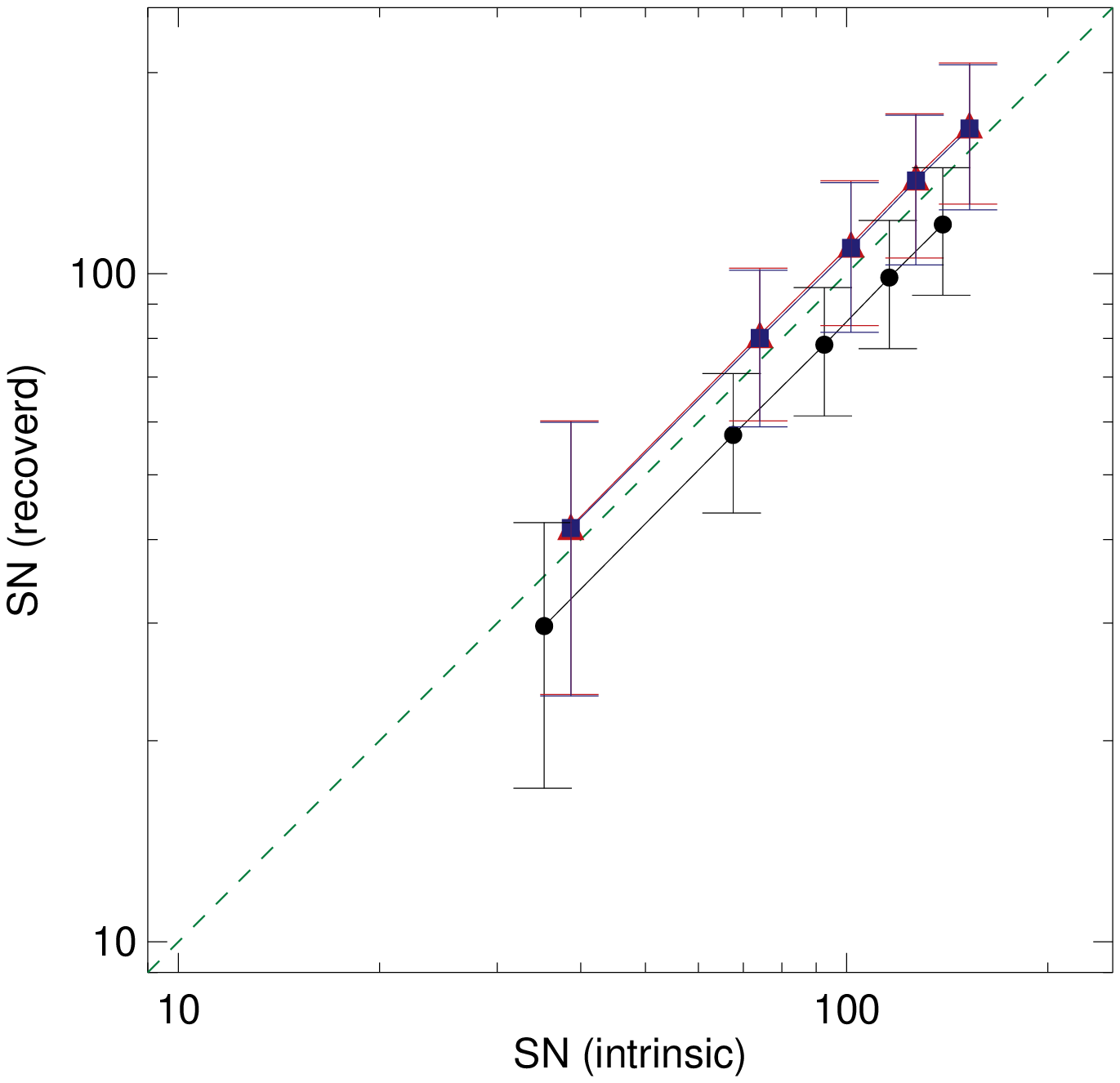}\\
\end{center}
\caption{\label{SN Errors all} Fig.~\ref{SN Errors AAOmega} is
repeated for each instrument configuration presented in
Table~\ref{Crosstalk modes} and with $\Delta$\,mag=0.  For the compact
fibre systems a significant systematic flux error is seen (for both
fibres) in the extracted flux due to double counting in the overlap
regions by the Tramline and Gaussian extraction methods.}
\end{figure*}

\begin{figure*}
\begin{center}
\includegraphics[width=7.5cm]{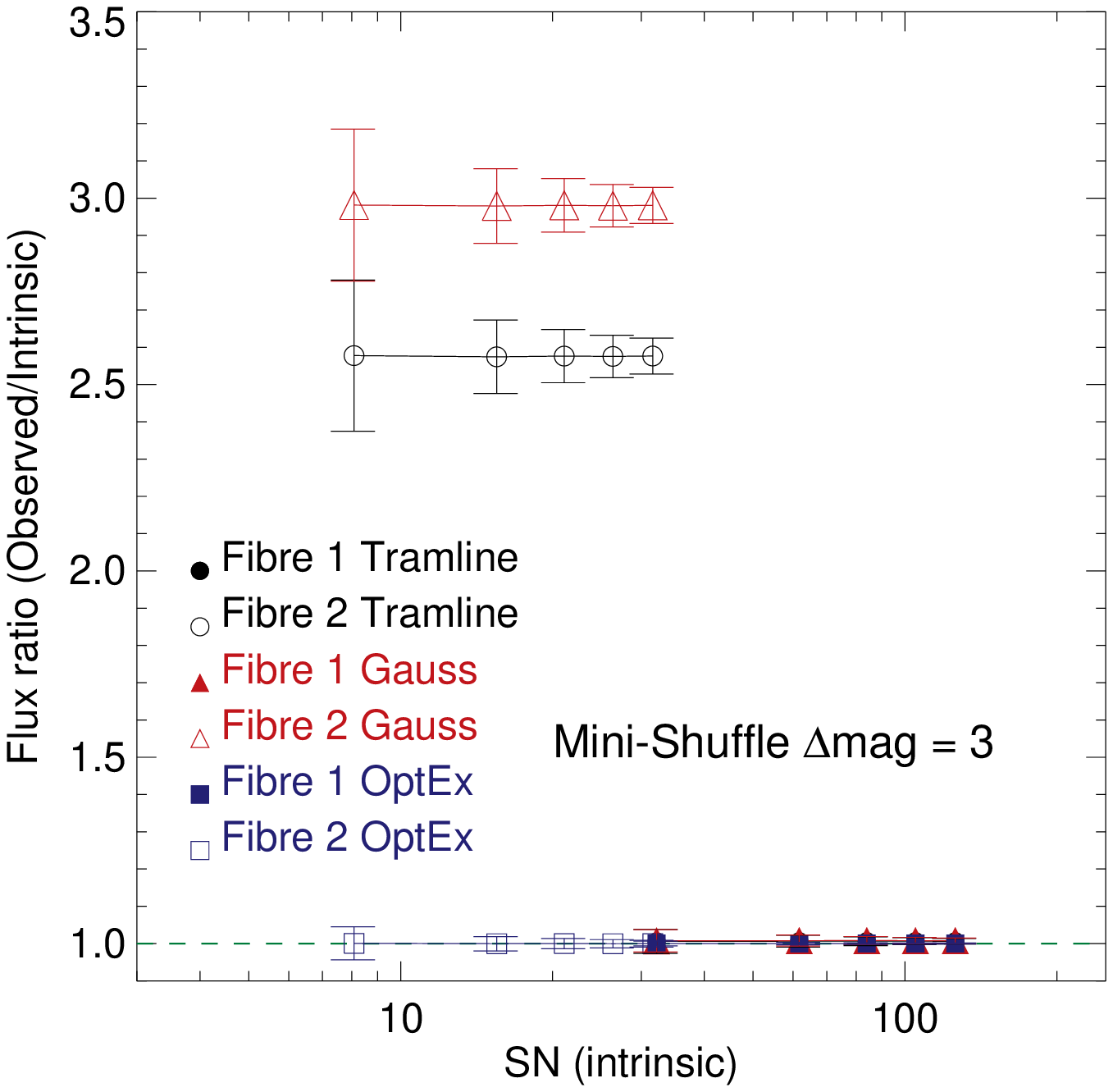}
\includegraphics[width=7.5cm]{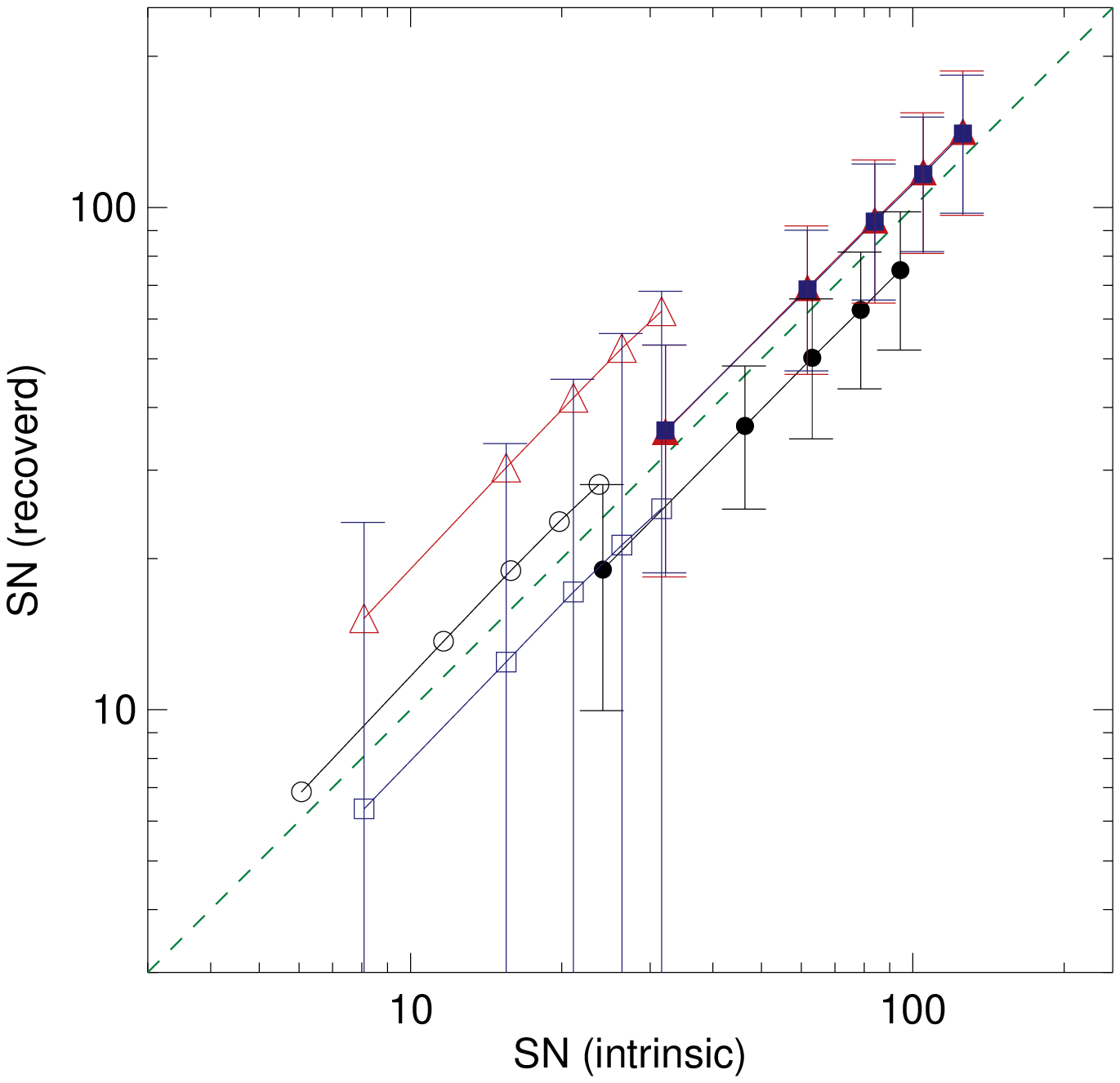}\\
\includegraphics[width=7.5cm]{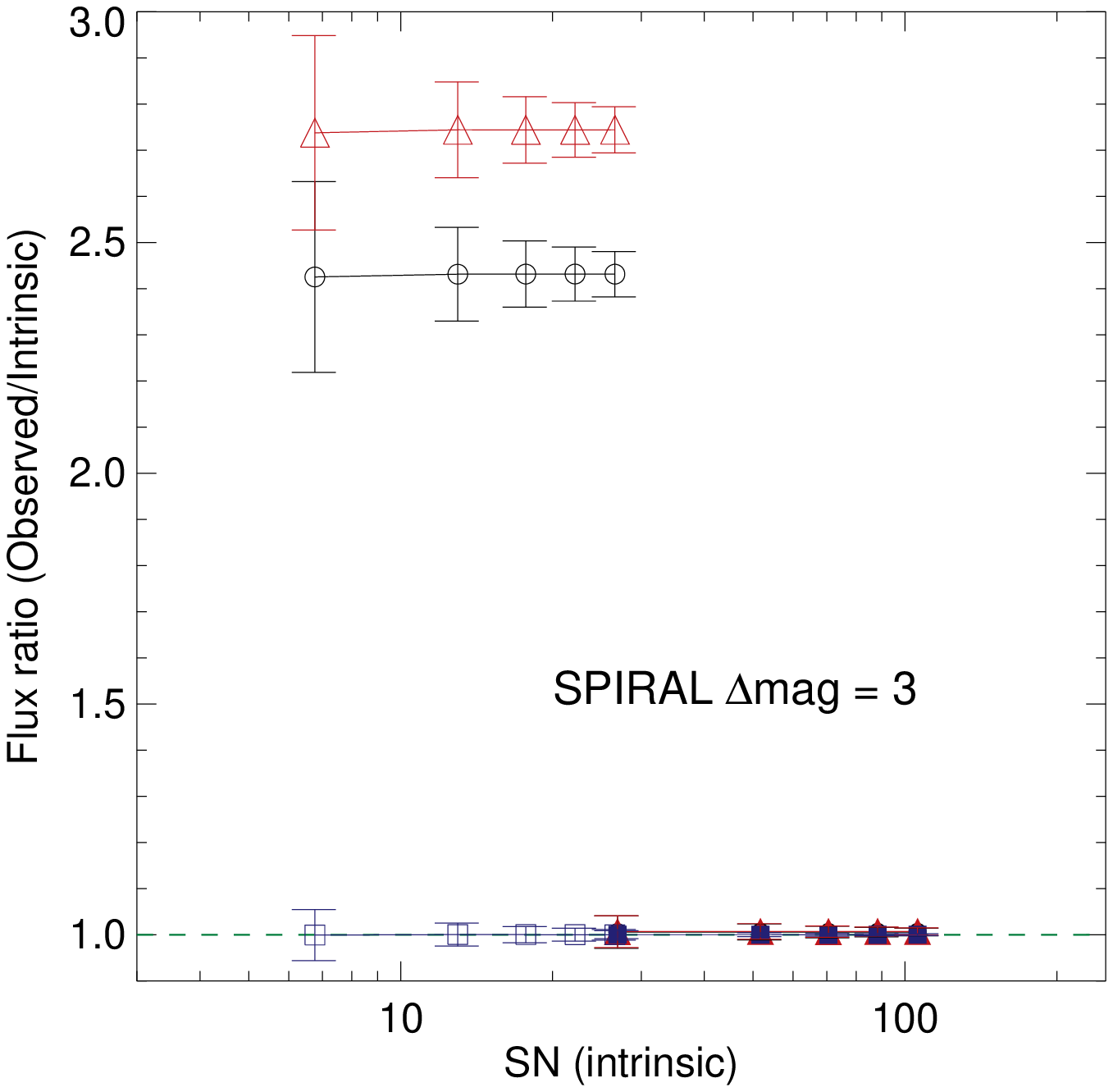}
\includegraphics[width=7.5cm]{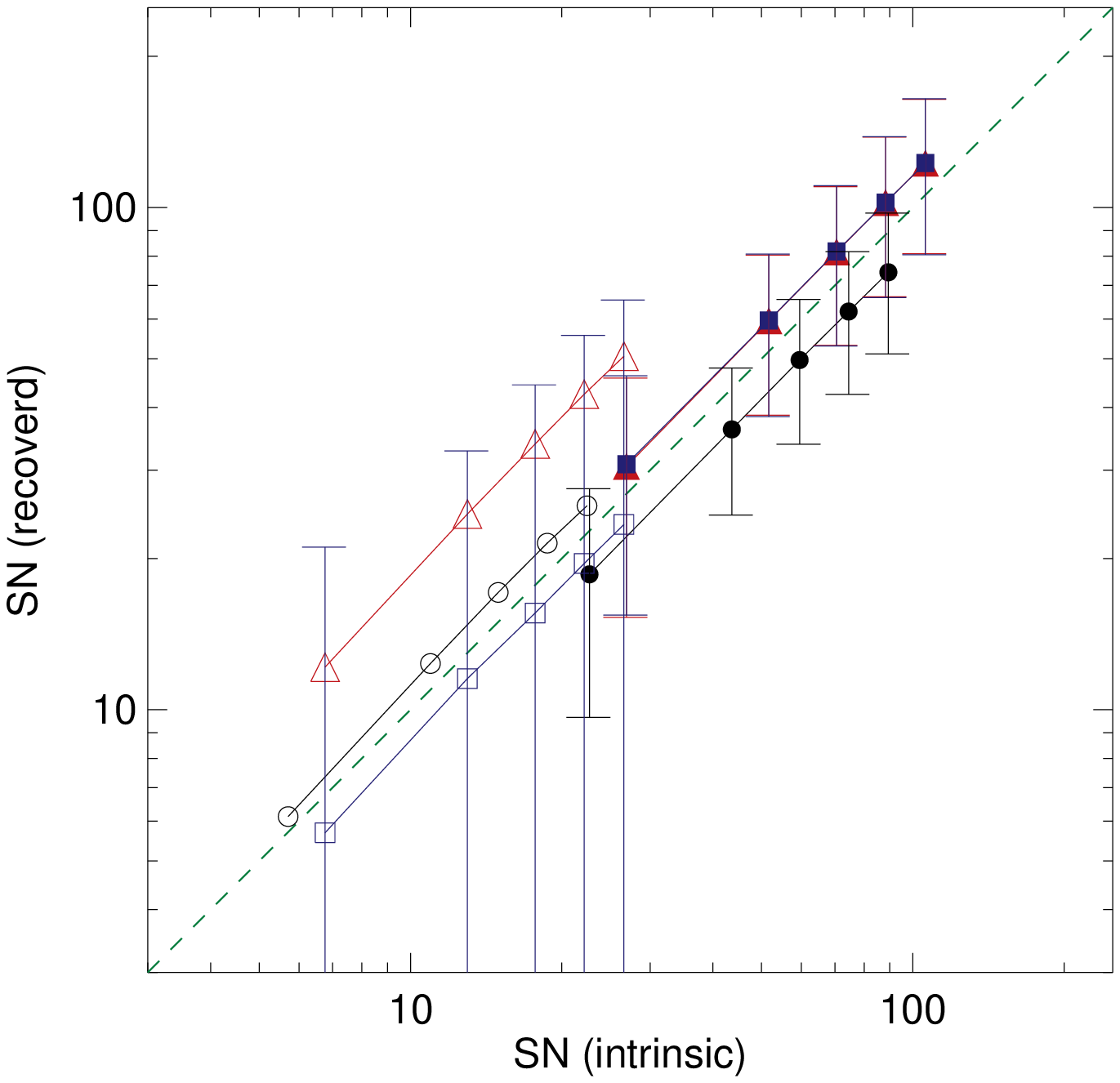}\\
\includegraphics[width=7.5cm]{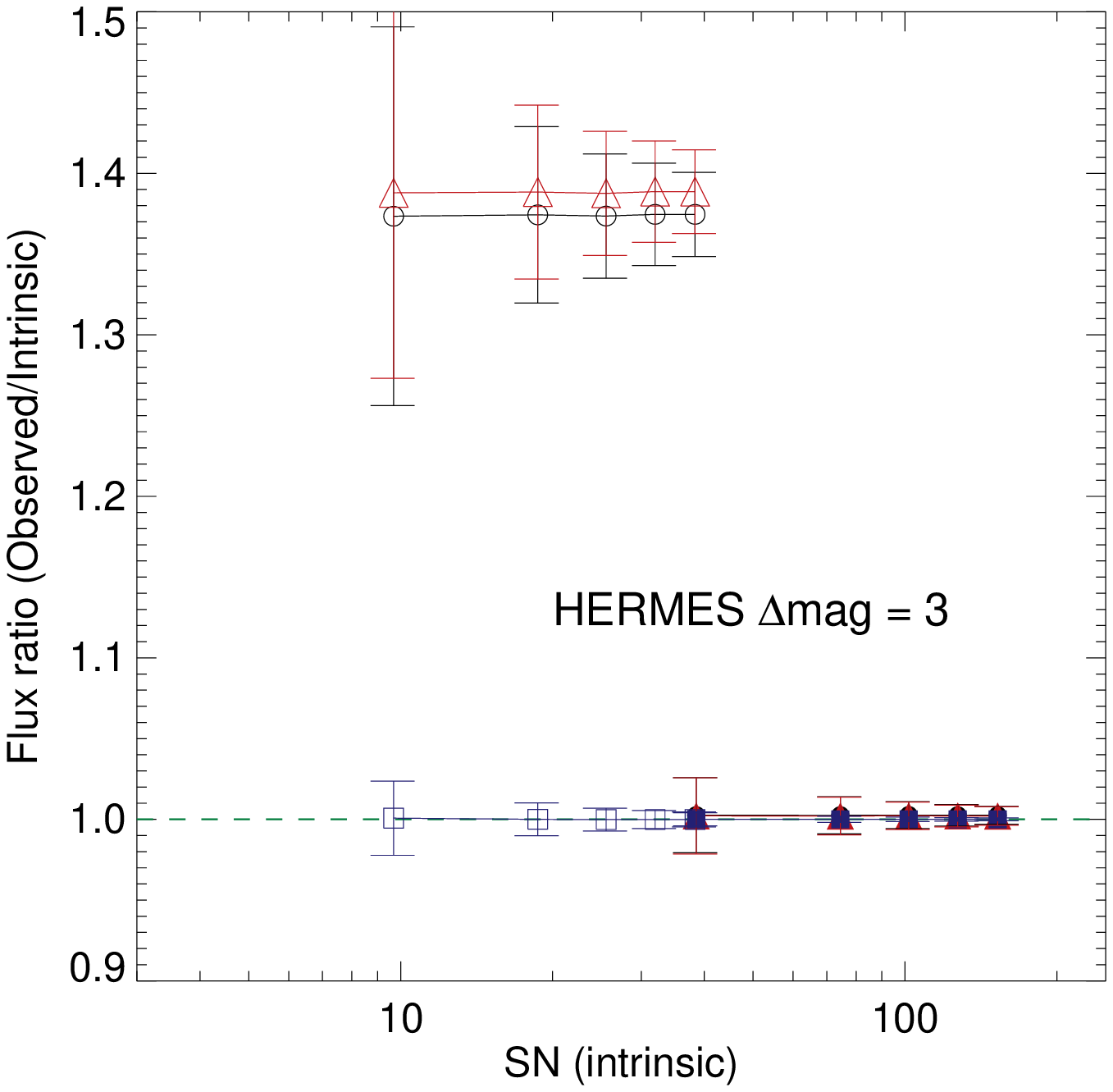}
\includegraphics[width=7.5cm]{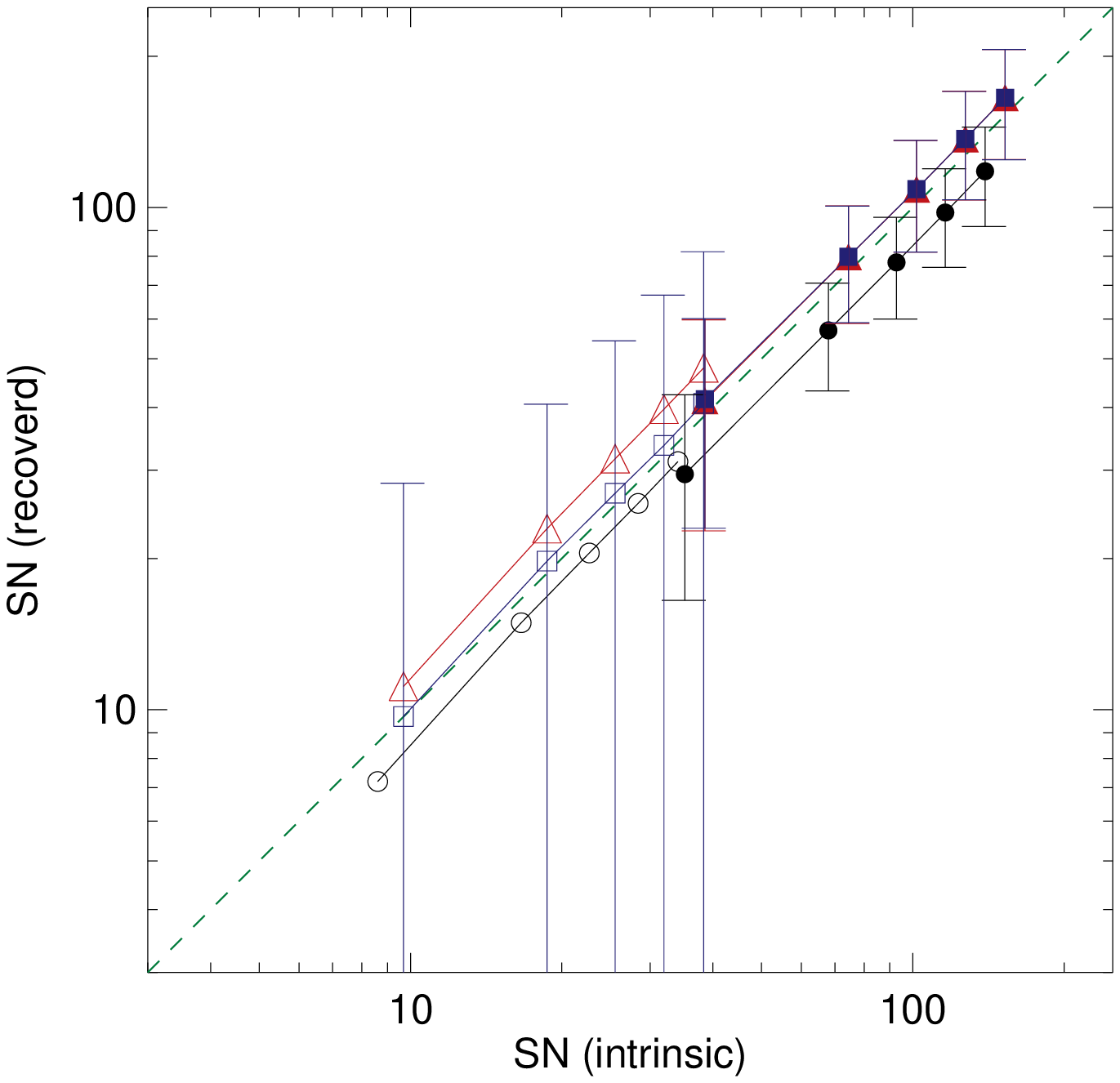}\\
\end{center}
\caption{\label{SN Errors all D3} Fig.~\ref{SN Errors all} is repeated
with $\Delta$\,mag=3.  Even with a large flux ratio between the fibres
the Optimal extraction method recovers the correct flux for both the
bright (fibre 1) and faint (fibre 2) profiles.  The apparent increase
in the SN ratio of the faint fibre with the tramline and Gaussian
extraction is due to the erroneous double counting of flux from the
adjacent fibres, which would represent a degradation of the SN ratio
actually achieved.}
\end{figure*}

\clearpage

\section{Profile extraction algorithms}
\label{algorithams}
In each of the three spectral extraction methodologies introduced in
\S\ref{spec ext} it is assumed that each fibre projects a spectrum
onto the detector such that the light is dispersed along rows of
pixels.  At each spectral pixel the fibre profile is assumed to be
well modelled by an analytic function aligned \emph{precisely} along
columns of the detector, orthogonal to the dispersion axis.  These
assumptions are approximately true at the pixel-to-pixel level, the
macroscopic spectral curvature on the CCD being on a much larger
scale.  Such an approximation is made in almost all approaches to the
extraction of spectra from 2D CCD data.

In the discussion of the two non-trivial spectral extraction
algorithms presented below we use the terms defined in
Table~\ref{symbols}.

\begin{table*}
\caption{\label{symbols} Definition of terms used in the discussion of
the extraction algorithms.  The desired products of the fitting
process are the integrated fibre intensities $\eta_k$ for each fibre
$k$, and the associated variance in $\eta_k$ expressed as $var_k$.
The process includes all pixels, $i$, along the spatial (slit) axis of
the CCD which is assumed orthogonal to the dispersion axis. It is then
repeated for each pixel of the dispersion axis.}
\begin{center}
\begin{tabular}{|cl|}
\hline
{\bf Parameter} & {\bf Definition}\\
\hline
\hline
%R & resudual\\
$N_{fib}$    & The number of fibre profiles in the system\\
$i$          & Running index over pixels in the spatial (CCD y-axis)\\
             & dimension orthogonal to the dispersion axis\\
$k$          & Running index over successive fibre profile 1-$N_{fib}$\\
$j$          & Dummy index for fibre $k$ when computing profile overlaps\\
\hline
\hline
$D_i$        & CCD data value at pixel $i$\\
$M_i$        & Model for data at pixel $i$\\
             & For small fibre pitch this is a sum over overlapping profiles\\
$\sigma_i$   & Error estimate at CCD pixel $i$\\
$Var(D_i)$   & Variance estimate at CCD pixel $i$ with data value $D_i$\\
\hline
\hline
$\eta_k$     & Integrated intensity for fibre $k$ at the current wavelength element\\
$var_k$      & Computed estimate of the variance associated with fibre $k$ with flux $\eta_k$\\
$\phi_i$     & Normalised fibre profile for a single fibre at spatial axis pixel $i$\\
$\phi_{ki}$  & Normalised fibre profile of fibre $k$ at pixel $i$\\
\hline
\hline
$c_{kj}$,$c'_{kj}$ & Internal variables for the cross terms between the profiles of fibre $k$\\ 
                   & and fibre $j$ for all fibres $k$,$j$=1-$N_{fib}$\\
$b_{j}$,$b'_{j}$   & Internal variables for the sum of products of the fibre profile $j$\\
                   & and the data/variance values at pixel $i$\\
\hline
\end{tabular}
\end{center}
\end{table*}

\subsection{Gaussian summation extraction via least squares}
\label{Gaussian extraction}
An extraction method which minimises the read-noise contribution from
pixels in the wings of the fibre profiles, while maintaining a full
summation of all of the observed flux, is in principle rather trivial
for the case of a single simple spectral profile (although careful
consideration of the system is required for a truly optimal result,
see \citet{horne86} for a comprehensive discussion).

We consider each column of the dispersed data in isolation, and
construct a model for the intensity at a given spatial pixel assuming
an appropriate fibre profile $\phi_i$, and integrated intensity
$\eta$.
\begin{equation}
M_i = \eta \times \phi_i\\
\end{equation}

In the case of non-overlapping spectra, and given a recorded CCD pixel
intensity D$_i$ with associated error estimate $\sigma_i$\footnote{The
error estimate for each data value is, as is typical for such data,
constructed from the data under assumptions of the detector read-noise
and gain characteristics and the \emph{shot-noise} from each data
value itself.  While this clearly leads to an imperfect estimate of
the error in the observation, we find this simple approximation to be
sufficient for our requirements.}, an estimate of the integrated
profile $\eta$ is given by the minimisation of\\
\begin{equation}
R = \sum_i\left( \frac{D_i - M_i}{\sigma_i} \right)^2
\end{equation}

This yields\\
\begin{equation}
\frac{dR}{dn_k} =
\sum_i\left( \frac{D_i \phi_i}{\sigma_i^2}\right)  - \sum_i\left( \frac{2 n_k \phi_i^2}{\sigma_i^2}\right)
\end{equation}
and on minimisation,
\begin{equation}
n_k = \sum_i\left(\frac{D_i \phi_i}{\sigma_i^2}\right) /
\sum_i\left(\frac{\phi_i^2}{\sigma_i^2}\right)
\end{equation}
with the summation running over a sufficiently large range of pixels,
$i$, to cover the full fibre profile, without overlapping with
adjacent profiles. A range of $\pm3\sigma$ (or half the inter-fibre
separation for tightly packed fibres) is typically adopted.  The
resulting value of the integrated profile flux $\eta$ is calculated
for each fibre in isolation and for each element along the dispersion
axis in tern.

\subsection{Multi-profile deconvolution extraction}
\label{the algoritham}
The core of the fibre profile achieved by the AAOmega system is
observed to be well matched to a Gaussian profile; and for speed and
simplicity, while still maintaining sufficient accuracy, the current
\texttt{2dfdr} data reduction environment used with AAOmega data
implements a simple Gaussian profile.

Furthermore, we assume that in the first instance the precise position
of each fibre profile as a function of wavelength (i.e.\ detector
column) is known from the previous trace of an appropriately
illuminated flat field frame as discussed in \S\ref{spec ext}.

We then define a normalised spatial profile for fibre $k$, as a
function of the CCD pixel $i$ along the spatial axis (the fibre slit
axis, orthogonal to the dispersion axis), such that
\begin{equation}
\sum_{i} \phi_{ki}=1
\end{equation}

Hence the contribution of fibre $k$ to the observed intensity value at
CCD pixel $i$ is given by the product of the integrated profile
intensity for the fibre, $\eta_{k}$, and the fibre profile at pixel
$i$.  For any given pixel of the detector a model value for the
recorded count rate at pixel $i$ is then simply the sum over all
contributing fibre profiles,

\begin{equation}
\label{model}
M_{i} = \sum_{k}\eta_{k}\phi_{ki},
\end{equation}

Given $D_{i}$ and $\sigma_{i}$, the recorded count rate and an
estimate of the statistical error at each pixel (along a single column
of the detector), we wish to evaluate $\eta_{k}$ for each spectrum.
This is achieved via the minimisation of the residual,

\begin{equation}
\label{R}
\mathrm{R}=\frac{1}{2}\sum_{i}\frac{\left(D_{i}-M_{i}\right)^2}{\sigma_{i}^{2}},
\end{equation}

\noindent and on setting $\frac{\delta\mathrm{R}}{\delta \eta_{k}}=0$
and substituting for $M_{i}$ we find,

\begin{equation}
  \sum_{k} \eta_{k} \sum_{i}
  \frac{\phi_{ji}\phi_{ki}}{\sigma_{i}^{2}}=
  \sum_{i}\frac{\phi_{ji}D_{i}}{\sigma_{i}^{2}},
\end{equation}

\noindent Letting $c_{kj}$=$\sum_{i}\frac{\phi_{ki}\phi_{ji}}{\sigma_{i}^{2}}$
and $b_{j}$=$\sum_{i}\frac{D_{i}\phi_{ji}}{\sigma_{i}^{2}}$\\
we find

\begin{equation}
\label{solve-me}
  \sum_{k} \eta_{k}c_{kj} = b_{j},
\end{equation}

Equ.~\ref{solve-me} can be solved using any of the multitude of
methods for solving coupled linear equations.  Numerical stability
considerations have guided our implementation towards a Singular Value
Decomposition (SVD) approach.  SVD allows terms of low significance to
be removed from the matrices during fitting, providing control over
numerical stability for high contrast data sets.
One notes that, even for very tightly packed fibre data, the overlap
between fibres separated by more than a few FWHM will be close to zero
in most instances (with the exception of very high contrast ratio
data, or for a model PSF with wide scattering wings).  This makes
$c_{kj}$ at least band diagonal, and tri-diagonal in many cases,
allowing trivial solutions to Equ.~\ref{solve-me} in these cases.

\subsubsection{Propagation of variance information}
\label{varprop1}
Propagation of error information is critical for all observational
science, but is woefully missing in many spectral extraction routines.
It can be achieved as follows.  For each pixel $i$ in the input image
we calculate the total of all fibre profiles, $\phi_{ki}$, that
contribute to that pixel,
\begin{equation}
\mathrm{T}_{i} = \sum_{k} \eta_{k} \phi_{ki}
\end{equation}
Then for each spectrum $k$, we calculate the fractional contribution
of this spectrum's profile to pixel $i$,
\begin{equation}
\mathrm{F}_{ki} = \frac{\eta_{k} \phi_{ki}}{\mathrm{T}_{i}}
\end{equation}
An estimate of the total variance in the extracted intensity $\eta_k$
is then given by
\begin{equation}
\mathrm{var}_{k}=\sum_{i} \mathrm{F}_{ki}^{2} \times \sigma_{i}^2
\end{equation}

\subsubsection{Alternate variance solution}
The variance estimate associated with any given detector pixel,
$Var(D_{i})$ is determined directly from the observed pixel value and
the predetermined read-noise and gain parameters for the CCD system.
Given Equ.~\ref{model} the value is given by the sum of the
contributions from each fibre at that pixel plus the additional
detector read-noise.  Under the assumption of shot-noise in the photon
arrival rate at a pixel, the contribution of a particular fibre to the
variance estimate for that pixel is given by the produce of the total
count rate for the fibre and the normalised profile intensity at the
pixel.
\begin{equation}
v_i = \phi_{ki} \eta_k = \phi_{ki} var_k
\end{equation}
When multiple fibre profiles contribute to a pixel, a model for the
observed variance in that pixel is then
\begin{equation}
M_i = \left(\sum_k \phi_{ki} var_k\right) + n_{rd}^2
\end{equation}
We then wish to minimise the residual between this model and the
variance in the observed data, Var($D_{i}$).

\begin{equation}
R=\frac{1}{2} \sum_i \left( Var(D_i) -M_i\right)^2,\\
\end{equation}
\begin{equation}
R=\frac{1}{2} \sum_i \left( \left(Var(D_i) - n_{rd}^2\right) - \sum_k \phi_{ki} var_k
\right)^2,\\
\end{equation}

By analogy with Section \ref{the algoritham}, and solving or for
$\delta$R/$\delta$$var_k$=0, we therefore arrive at an estimate for
the variance, $var_k$, associated with each integrated fibre profile
intensity $\eta_k$ by solving

\begin{equation}
\sum_{k} var_k c_{kj}' = b_{j}',
\end{equation}

with $c_{kj}'$=$\sum_{i} \left(\phi_{ki}\phi_{ji}\right)$ and
$b_{j}'$=$\sum_{i} \left(Var(D_{i}) -n_{rd}^2 \right) \phi_{ji}$.
This can be solved directly alongside Equ.~\ref{solve-me}, with
minimal additional computational overhead.

\subsubsection{Practical limitations and the iterative solution}
As demonstrated in Fig.~\ref{Errors}, variation in the fibre profile
FWHM with wavelength is of particular concern\footnote{The AAOmega red
camera suffers from a degraded ($\sim$0.2\,pixels) focus at long
wavelengths which is attributed to the fast (\emph{f/1.3)} camera and
increased photon penetration into the CCD before detection.}.  Poor
correction for background \emph{scattered light} will also prevent a
satisfactory solution being achieved.  Results can be significantly
improved via an iterative solution to the problem of determining these
free parameters.

We adopted an iterative solution with a number of free external
fitting parameters.  The model is derived from a high signal-to-noise
flat field frame and then the parameters are locked at the resulting
values when fitting for science frame intensities.  The free
parameters are\\

\vspace{-0.25cm}\noindent{\it Background -} We include a low order
polynomial fit to the broadband scattered light distribution across
the CCD (pedestal plus gradient).  In this model, scattered light is
effectively treated as a pedestal correction to the observed data.
For the AAOmega system, the scattering is well modelled as a DC
pedestal offset across the CCD.
This background level is a free parameter in the iterative solution of
Equ.~\ref{solve-me} (but see \S~\ref{non iter BG} for a further
refinement).\\

\vspace{-0.25cm}\noindent{\it Fibre Profile -} The normalised fibre
profile under the assumption of a Gaussian PSF has only a single free
parameter, $\sigma$, the Gaussian profile width.  A more accurate
accounting for the SPIRAL fibre PSF is considered in
\S~\ref{multi-component profile}.\\

\vspace{-0.25cm}\noindent{\it 2D parameter distribution -} A full 2D
polynomial fit to the variation of these free parameters as a function
of CCD position should also be performed.  This can be achieved by
fitting a subset of data columns at regular positions across the
dispersion axis.  A low order model would then be fitted for the full
data set guided by instrument design considerations.  This has not yet
been implemented within the \texttt{2dfdr} software.

With these external parameters in place we iteratively solve
Equ.~\ref{solve-me}, for all fibres simultaneously, for each element
of the dispersion axis in turn.  The \texttt{IDL} prototyping code
used
\texttt{MPFIT}\footnote{http://cow.physics.wisc.edu/$\sim$craigm/idl/fitting.html,
MPFIT, an excellent \texttt{IDL} implementation of the
Levenberg-Marquardt fitting procedure \citep{mpfit}} \citep{mpfit},
adopting the $\chi^2$ statistic as the figure of merit applied to the
difference between the data and model.

\subsubsection{Run time}
The multi-fibre deconvolution algorithm described above is
computationally intensive and to be of value the time taken to process
observational data must be considered.  The AAOmega system
(Table~\ref{Crosstalk modes}) requires the solution for $\sim$400
simultaneous fibre profiles ($\sim$800 for the mini-shuffle mode)
across 4000 spatial pixels for each of the 2000 spectral columns of a
2k$\times$4k E2V CCD.  The algorithm has been successfully implemented
in the \texttt{2dfdr} software environment (written primarily in
(\texttt{Fortran95}) and the extraction is completed for a single
frame in $\sim$1-2minutes on a modern desktop PC.

In its current format, fifty percent of the calculation time for each
instance of the fitting process is concerned with the calculation of
fibre profiles and the population of the $c_{jk}$ \& $b_j$ matrices
from Equ.~\ref{solve-me}.  The current implementation does not
distribute calculations across multiple PCUs and so this minor code
extension should allow for a near linear speed increase, for these
calculations, proportional to the a number of processors available on
a modern multi-core computer.

\section{The procedure}
\label{proc}
For data taken with the SPIRAL-AAOmega system we undertake the
following procedure

\begin{itemize}
\item
A fibre-flat-field frame is taken in which all the fibres are
illuminated with a uniform continuum source in order to trace the
centroid of the fibre profiles across the CCD.  Fibre-flat-field
frames can also be used to correct for variations between fibres in
the relative response function with wavelength.  They do not provide
an accurate correction for pixel-to-pixel CCD response variations due
to the high degree of spatial structure across the fibre profiles.  If
such a correction is required, e.g.\ to correct for interference
fringing in the CCD, a more uniform long-slit-flat field frame would
be required.
\item
All CCD frames are processed for overscan/bias correction and
population of the variance array information.
\item
An iterative fitting algorithm is applied to extract the integrated
fibre profile intensities for the flat field.  Experience with
\textsc{spiral} data shows that re-adjustment of the fibre centroids
measured from the fibre-flat-field is not required for AAOmega.
\item
The iterative fitting algorithm is applied to the science data,
holding the centroid and PSF values fixed (determined from the
fibre-flat-field above) and fitting only for the pedestal offset as a
free external parameter when iteratively solving for the integrated
fibre intensities via Equ.\ref{solve-me}.  Using the alternative
background subtraction method outlined in section \S\ref{non iter BG}
below, the extraction of science data can be made non-iterative once
the extraction parameters have been determined from the
fibre-flat-field data.
\end{itemize}

\section{Non-iterative background solution}
\label{non iter BG}
If a simple pedestal/polynomial model is adopted for the background
component then it is possible to fit the background without resorting
to an iterative solution.  For example, a simple pedestal can be
thought of as an additional fibre profile with a uniform profile
intensity at each pixel.

Higher order polynomial terms of the form $i^{N}$ could also be
introduced, up to some number of terms $N_{BG}$.  One would then solve
Equ.~\ref{solve-me} for $\eta_{k}$ with $1<k<N_{fib}$ for each fibre
with normalised fibre profile $\phi_{ki}$ and also for
$N_{fib}+1<k<N_{fib}+N_{BG}$ governing the background model intensity terms.

This more elegant formalism allows the solution of Equ.~\ref{solve-me}
without iteration in the case that all other external variables are
predefined.

\section{Scattered light}
\label{scattered light}
Two basic assumptions of the spectral extraction processes discussed
in this work are:\\

\vspace{-0.25cm}\noindent 1) The spatial (slit) axis on the CCD can be
considered orthogonal to the dispersion (spectral) axis at each pixel
along the dispersion axis\\

\vspace{-0.25cm}\noindent 2) Each spectral pixel is treated as if it
where independent of the adjacent spectral pixels, i.e.\ no accounting
is take of the 2D nature of the instrument PSF.\\

Essentially we are replacing the real-world spectrograph Point Spread
Function (PSF) with a pixel-by-pixel Line Spread Function
perpendicular to the dispersion axis.

Such assumptions are made primarily for computational expedience.  For
a compact and well-sampled PSF they are valid.  This is fortunate
since a full 2D modeling, which amounts to simultaneous deconvolution
involving every pixel on the CCD (and a number of virtual pixels
extending beyond the physical device), will likely remain beyond the
computing power available for routine multi-fibre observations for
some time due to the inherent scale of such a problem. As an example
of this scale consider that the 512 fibres of the AAOmega-SPIRAL
system each have $\sim$2000 spectral elements ($\sim$830 independent
resolution elements) covering $\sim$8\,million CCD pixels.

The model breaks down if the fibre PSF possesses a measurable
scattering wing which extends significantly beyond the core of the
fibre profiles.  If such a broad scattering profile is present then
the assumption that data can be accurately modeled as a 1D line spread
function rather than the 2D PSF breaks down since scattered light
present at any given position along the dispersion axis will be
largely dominated by the intensity of the local average spectral
intensity of the surrounding spectrum and unrelated to the specific
spectral intensity of the current spectral pixel.

In high signal-to-noise data (including the fibre-flat-field frames
for science data that will ultimately be in the low signal-to-noise
regime), even a scattering fibre profile with low overall percentage
light level in the scattering component can have significant side
effects on the ultimate data quality if the profile has a broad
PSF. The scattered light from an increasing number of more remote
fibre profiles will become more and more important.

\subsection{Scattering without fibre structure}
Some scattered light is inevitable in all optical systems.  Low level
background structure which is largely devoid of the high spatial
frequency signature of the fibres themselves can be removed relatively
easily.  One may fit directly to the background level observed in
regions free from {\it contamination} by the fibre profiles.
Alternatively a low order model can be included as an additional {\it
fibre} profile as demonstrated in \S\ref{non iter BG}.  This model
profile would ideally be informed by an independent modeling of the
background scattered light component.

\subsection{Multi-component profile model}
\label{multi-component profile}
For scattering more directly associated with the extended wings of the
individual fibre PSFs, a more complex consideration is required.  As
previously mentioned, a full 2D deconvolution across the entire CCD
array is likely beyond the capabilities of modern systems, at least at
a rate comparable with the data rate from modern multi-fibre
spectrographs.

A solution is to maintain the 1D orthogonal extraction assumption, but
using fibre profiles with extended wings.  This approach is flawed
however. For an extreme example of why consider the case of a strong
stellar absorption line within a strong continuum source.  The
scattered light at line centre would be dominated by the strong local
continuum average, due to the underlying 2D nature of the scattered
light PSF.  The absorption line core profile would however be rather
weak in comparison.  This local variation in the spectral shape
essentially induces a local variation in the assumed line spread
function.  One cannot merely fit for a core profile with a scattering
wing modeled as a fixed percentage of the core profile flux.

A solution is to model each fibre profile as a composite of two
normalised profiles, one for the core profile, $\phi^c_i$, and a
second broader scattering profile $\phi^s_i$.  Each component will
have an independent intensity, $\eta^c$ and $\eta^s$. The model for
the CCD pixel intensity at pixel $i$ is then the sum, over all fibre
profiles $k$ which contribute to the pixel, of both the core and
scattering profiles.
\begin{equation}
M_i=\sum_{1<k<N_{fib}} \left(\eta^c_k \phi^c_{ki} + \eta^s_k \phi^s_{ki}\right),
\end{equation}
Equ.~\ref{solve-me} is now solved for 2$\times$$N_{fib}$ fibre profile
intensities $\eta_k$.

There are inherent degeneracies in this composite profile description
for each fibre.  At low signal levels one can encounter significant
problems with numerical stability when solving for values of $\eta_k$
in real data.  We have found the best results are achieved via a
Single Value Decomposition (SVD) approach.  The opportunity to
suppress poorly behaved elements of the intermediate solution matrices
within the SVD solution (akin to discarding from the analysis the
coupled linear equations which contain limited information, (as
discussed by \citet{NumRep}) recovers numerical stability and more
than compensates for the increased computational burden presented by
SVD over simpler solvers.  The SPIRAL IFU data extraction shown in
Fig.~\ref{2DIFU} were achieved using a composite profile extraction
and an SVD solver.

\subsection{Scattering from beyond the free spectral range}
An interesting side-effect is encountered as one approaches the edges
of the CCD.  For a spectrograph with a flat response function in
wavelength, scattering into an individual fibre profile from shorter
and longer wavelengths will be much the same as that scattered out of
the element in question (provided the spectral shape does not possess
strong spectral breaks).  However, as one approaches the limit of the
free spectral range of the spectrograph this symmetry no longer
holds. The exigencies of spectrograph design will typically place
sharp spectral features such as those from order sorting filters,
dichroic beam splitters or the grating blaze {\it just beyond} the
operational range of the spectrograph.  This means that the scattered
light arriving in pixels towards the edge of the detector array may
manifest a significant reduction in intensity since little light is
scattered back onto the detector from {\it just beyond} the observed
wavelength range.  This has been found to be particularly true for the
AAOmega system since the Volume Phase Holographic (VPH) gratings it
employs have strong gradients in the blaze profile at the extreme ends
of the intended observational ranges.

\section{Discussion}
\label{discussion}
A striking comparison of fibre-to-fibre crosstalk using the {\it
tramline}, {\it Gaussian} and {\it optimal} extraction techniques is
given in Fig.~\ref{2DIFU}.

The left panel of Fig.~\ref{2DIFU} shows the reconstructed image
obtained using the simple \emph{tramline} extraction model.  The image
is created by collapsing a 3D SPIRAL data cube over the spectral
dimension, to create a low resolution (0.7\,arcsec pixels) image of a
standard star observation.  The vertical structure is not a telescope
diffraction spike, but rather the manifestation of fibre-to-fibre
crosstalk.  The SPIRAL fibres are arranged in 16 banks (IFU short
axis) of 32 fibres (IFU long axis).  Hence all fibres in the central
column of the IFU are subject to scattered light in the wings of the
profiles of the fibres close to the bright star under observation.
With a simple \emph{tramline} extraction, which fails to account for
this crosstalk, the central columns of the IFU exhibit an enhanced
light level in all spectra in the central columns.  For more exotic
astronomical objects, any and all columns of the IFU with high
contrast data (either in the form of intensity or spectral type
variations) will suffer a similar fate.

For the central pane of Fig.~\ref{2DIFU} the extraction was performed
using the {\it Gaussian weighting}.  While this methods optimises the
signal-to-noise ratio for each spectrum, it fails to account for the
cross-contamination between the spectra.

The right most image shows the same data after applying the {\it
optimal extraction} technique.  A round stellar image is finally
achieved, with an order of magnitude reduction in any signature of
fibre-to-fibre cross contamination.

\begin{figure*}
\begin{center}
\includegraphics[width=5.2cm]{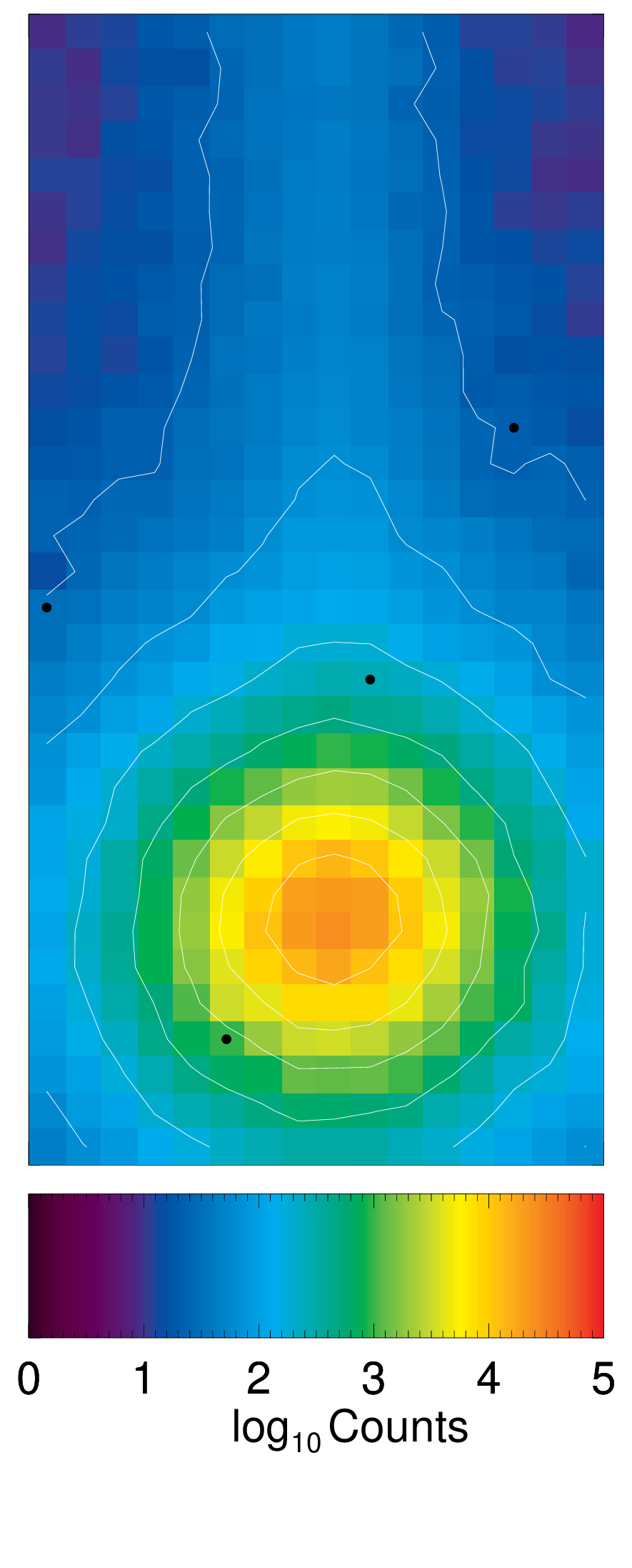}
\includegraphics[width=5.2cm]{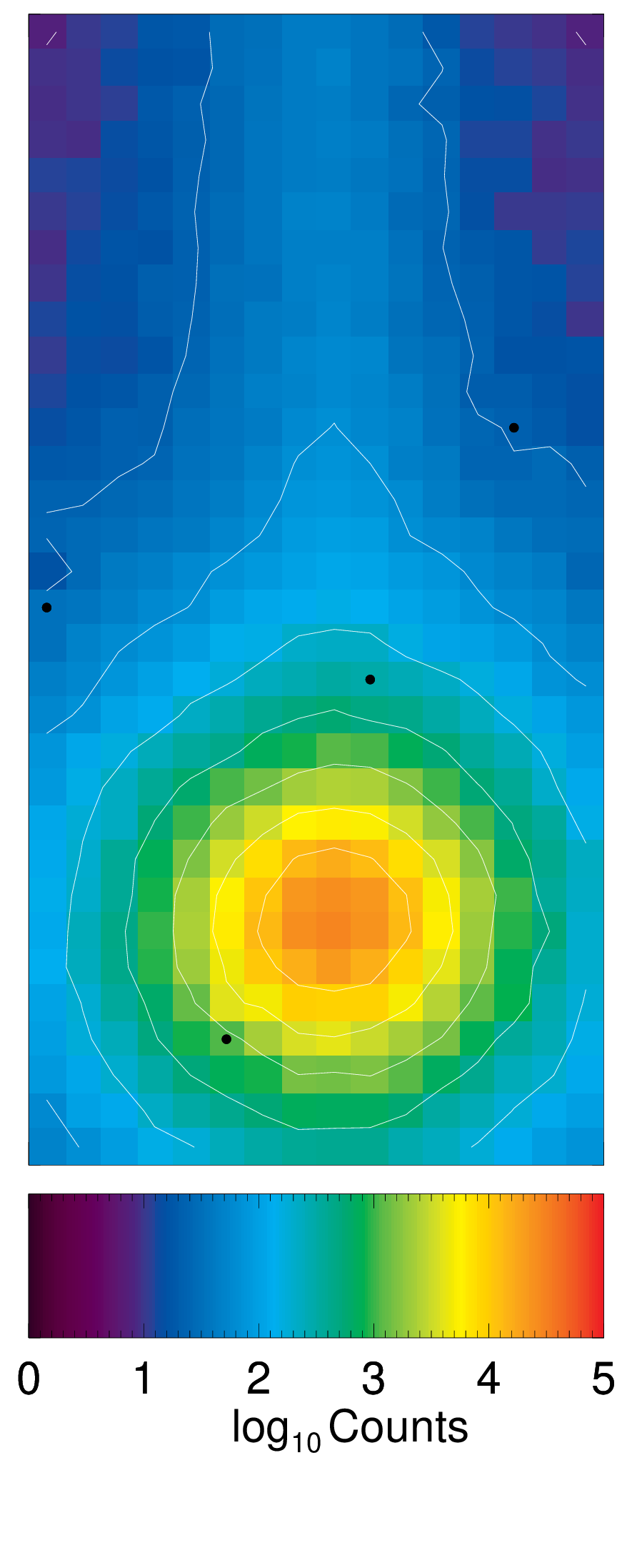}
\includegraphics[width=5.2cm]{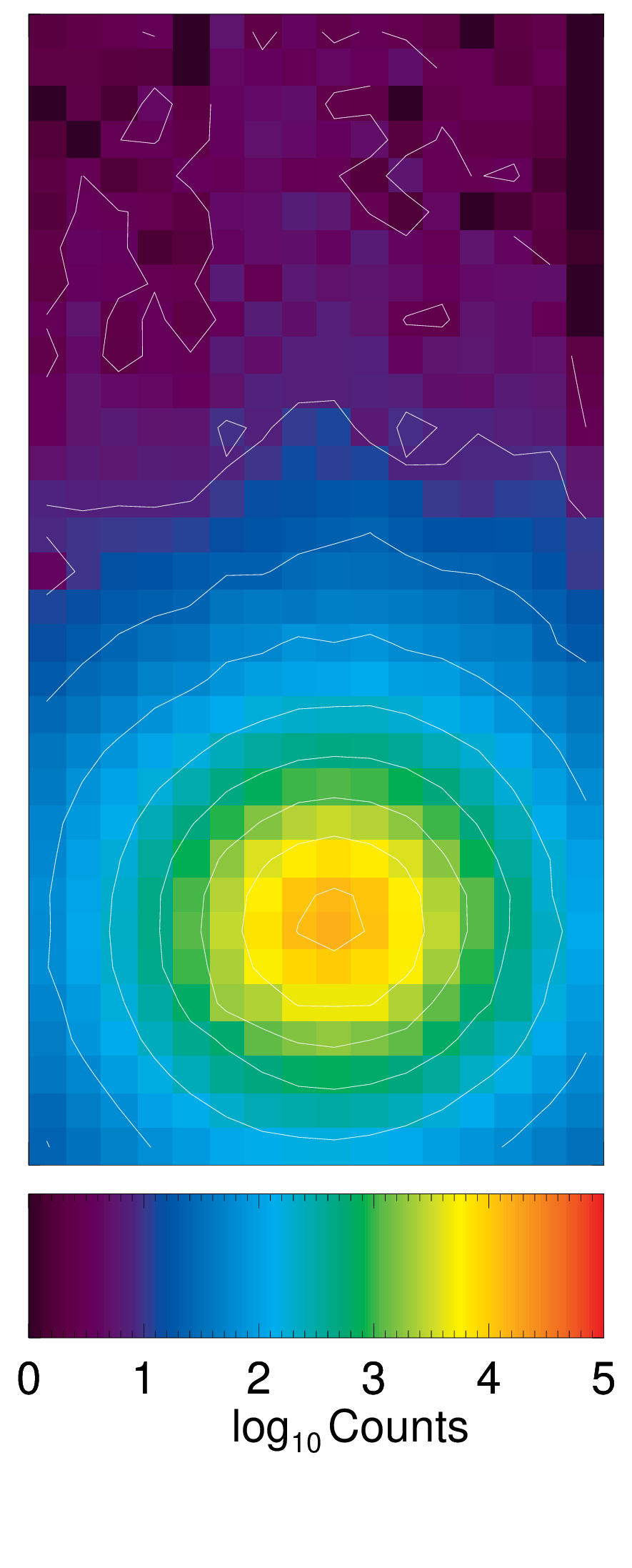}
\end{center}
\vspace{-1.5cm}
\caption{\label{2DIFU} A reconstructed 2D image, created by collapsing
the spectral data cube over wavelength, is shown for a SPIRAL-IFU
observation of a standard star.  Four {\it dead fibres} have been
interpolated in the data, and are marked with small spots.
\newline
Left - The reconstructed image obtained using a simple
\emph{tramline} extraction model.  The vertical structure is not a
telescope diffraction spike, but rather an artifact of fibre-to-fibre
crosstalk.
\newline
Centre - The {\it Gaussian summation} extraction suppresses noise
but not the fibre-to-fibre cross contamination.
\newline
Right - The reconstructed image obtained using the \emph{optimal}
extraction model.  The vertical extraction artifact structure removed
and a rounded PSF is recovered.}
\end{figure*}

\section{Conclusion}
We have presented an optimal extraction methodology for use with
multi-fibre spectroscopy in the regime where fibres are tightly packed
onto the detector array.  We demonstrate that high accuracy can be
achieved with this approach, minimising the impact of fibre-to-fibre
cross-contamination.  Additionally we show that the Gaussian least
squares fitting extraction used by default for AAOmega-MOS data reduce
within \texttt{2dfdr} is adequate for the well-separated fibre
profiles of this instrument mode provided the range of input target
magnitudes is kept below $\Delta$\,mag$<$3.

\section*{Acknowledgments} %If needed
The {\it optimal extraction} approach presented in this paper is based
on the extraction process developed by Rachel Johnson and Andrew Dean
for the \textsc{cirpass} instrument.  We thank Will Saunders and Scott
Croom for helpful discussions on the process, and the anonymous
referee for insightful comments.

\end{document}